\documentclass{article}

\usepackage{arxiv}

\usepackage[utf8]{inputenc} % allow utf-8 input
\usepackage[T1]{fontenc}    % use 8-bit T1 fonts
\usepackage[hidelinks]{hyperref}       % hyperlinks
\usepackage{url}            % simple URL typesetting
\usepackage{booktabs}       % professional-quality tables
\usepackage{amsfonts}       % blackboard math symbols
\usepackage{amsmath}
\usepackage{nicefrac}       % compact symbols for 1/2, etc.
\usepackage{microtype}      % microtypography
\usepackage{graphicx}
\usepackage{natbib}
\usepackage{doi}
\usepackage{subcaption}

\title{A deep neural network physics-based reduced order model for dynamic stall}

\date{} 					% Or removing it

\author{
	\href{https://orcid.org/0000-0002-6664-4170}{\includegraphics[scale=0.06]{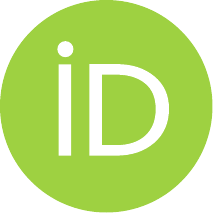}\hspace{1mm}Giacomo Baldan} \\
	Department of Aerospace Science and Technology \\
	Politecnico di Milano\\
	Milano, 20156 \\
	\texttt{giacomo.baldan@polimi.it} \\
	\And
	\href{https://orcid.org/0000-0001-6432-2461}{\includegraphics[scale=0.06]{orcid.pdf}\hspace{1mm}Alberto Guardone} \\
	Department of Aerospace Science and Technology \\
	Politecnico di Milano\\
	Milano, 20156 \\
	\texttt{alberto.guardone@polimi.it} \\
}

\renewcommand{\headeright}{}
\renewcommand{\undertitle}{}
\renewcommand{\shorttitle}{A deep neural network physics-based reduced order model for dynamic stall}

\hypersetup{
pdftitle={A deep neural network physics-based reduced order model for dynamic stall},
pdfauthor={Giacomo Baldan, Alberto Guardone},
}

\begin{document}
\maketitle

\begin{abstract}
	A physics-based machine learning framework is developed to compute the aerodynamic forces and moment for a pitching NACA0012 airfoil incurring in light and deep dynamic stall.
	Three deep neural network frameworks of increasing complexity are investigated: two multilayer perceptrons and a convolutional neural network.
	The convolutional framework, in addition to the standard mean squared error loss, features a physically-informed improved loss function to compute the airfoil loads.
	In total, four models are investigated of increasingly complexity.
	The convolutional model, coupled with the physics-based loss function, is found to robustly and efficiently predict pressure and skin friction distributions over the airfoil over the entire pitching cycle.
	Periodic conditions are implemented to grant the physical smoothness of the model output both in space and time.
	An analysis of the training dataset point distributions is performed to point out the effects of adopting low discrepancy sequences, such as Latin hypercube, Sobol’, and Halton, compared to random and uniform sequences.
	The current model shows unprecedented performances in predicting forces and pitching moment in a broad range of operating conditions.
\end{abstract}

% keywords can be removed

\section{Introduction}
Dynamic stall can be encountered in a wide range of aeronautical applications, such as over blades of helicopter rotors and maneuvering fixed wing aircraft. It is of concern also for wind turbines and turbomachinery applications.
Dynamic stall is a highly nonlinear unsteady aerodynamic phenomenon observed at large angles of attack, in combination with rapid variations of incidence. 

The ability to predict and mitigate dynamic stall results in an improvement of safety standards and also a better estimation of the flight envelopes~\citep{Gardner2023}. The occurrence of dynamic stall is also one of the critical factor limiting the maximum speed of conventional helicopters.
Compared to static stall, dynamic stall develops at a larger effective angle due to the rapid increase of incidence that temporary delays the boundary layer separation.
The delay in the boundary layer separation is attributed to two effects. 
One is an increase in the effective camber that is predicted by quasi-steady thin airfoil theory. 
The other is the acceleration of the boundary layer due to the Magnus effect produced by the motion of the leading edge~\citep{Corke2015}.
The boundary layer separation is followed by a large flow detachment on the suction side of the profile. 
This results in a disrupt loss of lift, a strong pitch down moment, and an increase in drag. 
Before flow reattachment, varying loads are observed due to the chaotic nature of the involved phenomena~\citep{Gardner2019}. 
The variation of the effective angle of attack can be generated by a blade pitch, like in helicopter blades, or due to a change in the relative flow velocity, as in gust encounters~\citep{Smith2020,Corke2015}.

Many experiments and computational simulations have been performed to investigate dynamic stall phenomena and study the flow around pitching and plunging airfoils~\citep{Khalifa2023,Zanotti2014a}. 
Especially during the last decade, thanks to the advances in numerical methods and computational power provided by modern GPGPUs~\citep{Devanna2022, Devanna2023} together with optical measurements techniques~\citep{Damiola2023b,Zanotti2013,Zanotti2014b}, dynamic stall has received particular attention in the literature where considerable advances in modeling, prediction, and understanding are reported.
Several aspects of dynamic stall have been numerically investigated ranging from aspect ratio~\citep{Hammer2021, Hammer2022} to sweep angle~\citep{Hammer2023} and compressibility effects~\citep{Benton2020}. The current state-of-the-art numerical simulations consist in wall-resolved large eddy simulations (LES) of ramp-up motion for spanwise-extruded airfoils~\citep{Visbal2018, Miotto2022}.

One key aspect, relevant to many practical applications, is the formulation of a reduced order model (ROM) for dynamic stall. 
Due to the large computational costs associated to dynamic stall CFD simulations, ROM are needed to reduce the computation cost of \textit{e.g.} aeroelastic analysis or flight mechanics evaluations. 

The two most common models nowadays are Leishman-Beddoes and ONERA ones.
Beddoes model was originally developed in the 1970s in indicial formulation~\citep{Beddoes1976, Beddoes1978}. 
Then, further improvements led to the well-known Leishman-Beddoes model~\citep{Leishman1986, Leishman1988, Leishman1989b}.
For aeroelastic applications, a state-space representation is available in the literature~\citep{Leishman1989a,Leishman1990}.
The latest version of the Leismann-Beddoes model is reported by ~\citet{Beddoes1993} and it is known as the \textit{Third Generation Model}. This model improves the accuracy of predicted loads for low Mach numbers and low reduced frequencies, by adding two equations with respect to the previous generation model.
The ONERA model describes the unsteady airfoil behavior using a set of nonlinear differential equations, to predict the response in the time domain~\citep{Tran1981a, Tran1981b}.
In contrast to the Leishman-Beddoes one, the ONERA formulation does not include physics-based modeling but instead focuses on the hysteresis behavior.
\citet{Truong1993, Truong1996} presents the latest version of the model, called \textit{ONERA BH model}.

Recent developments in dynamic stall ROM leverage on high-fidelity numerical simulations to better understand the underlying physics of dynamic stall and gather data from larger datasets.
The two most common algebraic methods present in literature to extract spatio-temporal flow features, are the Proper Orthogonal Decomposition~(POD)~\citep{Lumley1967} and the Dynamic Mode Decomposition~(DMD)~\citep{Rowley2009, Schmid2010}.
An example of a dynamic stall ROM, which uses Spectral Proper Orthogonal Decomposition (SPOD) to decompose the velocity flow field from a Delayed Detached Eddy Simulations (DDES) with SST model, has been developed in Refs.~\citet{Avanzi2021, Avanzi2022}.
The approach by \citet{Avanzi2021, Avanzi2022} defines an appropriate operating range of the filter’s typical dimension to identify coherent structures in the process and reconstructs force and moment coefficients over the pitching cycle using a reduced set of modes.
A recent extensive work in the same direction is presented by~\citet{Chiu2023}, where large eddy simulations (LES) with SPOD are adopted to analyze the impact of coherent structures on the aerodynamic forces.
The authors discovered that the zero-th frequency of the first mode is linked to an oscillating near-wall stream that adheres to the reattachment flow pattern. 
The first frequency is associated with a pair of counter-rotating vortices emerging at the point of flow reattachment. 
Finally, the second frequency of the first mode corresponds to smaller counter-rotating vortex pairs at the shear layer, originating in close proximity to the reattachment point.
\citet{Taha2021} and \citet{Olea2022} present an extension of the classical airfoil theory to high angles of attack. 
The model is designed to capture unsteady nonlinear characteristics at high angles of attack, and to conveniently apply nonlinear system analysis tools such as geometric-control theory and averaging.
\citet{Tatar2020} derived a nonlinear reduced model of the dynamic stall using a fuzzy inference system (FIS) and the adaptive network-based FIS (ANFIS). 
In addition, the Gram–Schmidt orthogonalization technique is used to construct a high-order set of the input variables to improve the prediction at post-stall angles of attack.
The proposed reduced-order model accurately predicts the aerodynamic response of the pitching airfoil at two reduced frequencies.
\citet{Glaz2013} describe a surrogate-based recurrence framework approach based on non-stationary Gaussian process models as a promising alternative to widely used semi-empirical rotorcraft dynamic stall models.

A powerful tool that is becoming more and more popular in engineering applications is machine learning \cite{Brunton2020}.
Recent developments in deep learning bring advanced and innovative approaches to improve the efficiency, flexibility, and accuracy of the predictive models~\citep{Vinuesa2023, Chen2023}. 
Some of the outstanding applications of deep neural networks (DNNs) in the domain of computational physics are solution of partial differential equations (PDEs), like Physics-Informed Neural Networks (PINNs)~\citep{Eivazi2022a, Bragone2022a, Bragone2022b, Baldanm2021, Baldanm2023}.
Other neural network architectures that allow to analyze information from large datasets are Convolutional Neural Networks (CNNs).
They received increasing attention by the fluid-mechanics community for their ability in pattern recognition~\citep{Guastoni2021, Eivazi2022b}.
In addition, convolution layers are usually coupled with pooling and up-sampling layers, allowing to reduce or increase data size, respectively.
The neural network capability of analyzing a large dataset has been of particular interest in dynamic stall phenomena since time-dependent series can be analyzed as a whole.
\citet{Lennie2020} use a clustering method based on dynamic time warping to identify different vortex shedding behaviors in experiments.
The work analyses cycle-to-cycle variations and on procedures to to detect outliers in separated flows.
A convolutional neural network is used to extract dynamic stall vorticity convection speeds and phases from pressure data.
The method applies to both static and dynamic stall and captures the occasional reattachment of the flow under surging flow in static stall and the strong variation of secondary and tertiary vorticity vary strongly in dynamic stall.
A further example of machine learning applied to dynamic stall can be found in~\citet{Kasmaiee2023}. 
A multi-layer perceptrons (MLP) neural networks is used to train the aerodynamic coefficients as functions of the control parameters and reduce the number of simulations.
A genetic algorithm is employed to optimize the configuration of a pure suction jet actuator on an oscillating airfoil.
~\citet{Damiola2023a} proposed a State-Space Neural Network (SS-NN) model trained using sine sweep functions at several angle-of-attack ranges. 
The study shows that SS-NN can be a powerful tool to accurately predict the unsteady aerodynamic loads of a pitching airfoil, both in pre-stall and post-stall conditions. 
In particular, the model succeeds in correctly capturing highly nonlinear flow features such as the delay of flow separation, and the formation and shedding of the dynamic stall vortex.
A final example is presented by~\citet{Eivazi2020}, where an autoencoder network is used for non-linear dimension reduction and feature extraction as an alternative to singular value decomposition (SVD). 
Then, the extracted features are used as an input for a long short-term memory (LSTM) network to predict the velocity field at future time instances. 
~\citet{Solera2023, Wang2024} further confirmed the NN capabilities of predicting dynamic stall loads by comparison with non-intrusive reduced order models based on dynamic mode decomposition and proper orthogonal decomposition. 

This work presents a deep neural network-based framework designed to predict force and moment coefficients throughout an entire pitching cycle, encompassing both light and deep dynamic stall conditions. 
The primary goal is to define a reduce order model that can efficiently predict the aerodynamic loads for integration into an aeroelastic code. 
Three deep neural networks of increasing complexity are applied: two  multilayer perceptrons and a convolutional neural network.
The operating conditions are defined by three parameters: the amplitude of the pitching angle $\alpha_s$, the reduced frequency $k$, and the free-stream velocity $V_\infty$.
In section~\ref{sec_dataset_generation}, the dataset point sequences are reported with the numerical setup of computational fluid dynamics simulations to generate the datasets.
In section~\ref{sec_neural_network_architectures}, the three neural network frameworks are presented together with the physics-based loss function for the convolutional framework. 
After describing the numerical setup, the performances of the four models are investigated in detail in the section ~\ref{sec_results}, highlighting the pros and cons.
Finally, section~\ref{sec_conclusions} reports the conclusions and the final remarks.

\section{Aerodynamic dataset generation} \label{sec_dataset_generation}

\begin{figure}
	\centering\includegraphics[width=0.6\textwidth]{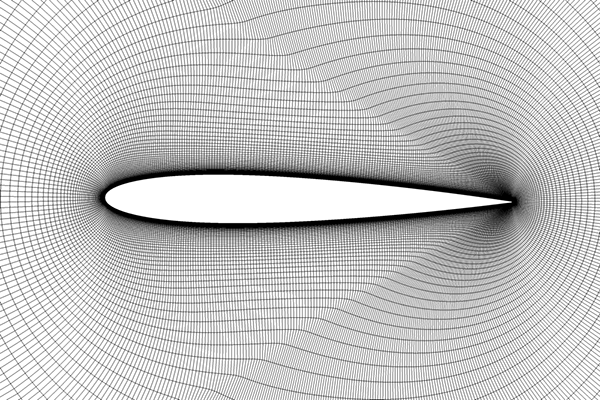}\\
	\caption{O-grid detail used in the computations.}
	\label{img_ogrid}
\end{figure}

The aerodynamic dataset for training the model is generated by solving the unsteady 2D RANS equations in selected conditions.
The numerical setup is assessed against the experimental campaign presented in~\citet{Lee2004}. 
A detailed grid convergence analysis and a study of the effects of the turbulence model is reported by~\citet{Baldan2023}.
The main settings of the reference test case are summarized in the following for completeness.
The reference case is the NACA 0012 airfoil with a chord $c$ of 0.15 m, undergoing a sinusoidal pitching motion, at reduced frequency k~=~0.1 and Reynolds number Re~=~$1.35\cdot10^5$.
The free-stream velocity is $V_\infty$~=~14~m/s with a turbulent intensity equal to 0.08\%, pressure is $P_\infty$~=~1~atm, and the pitching frequency $\omega$ is set equal to 18.67~Hz. 
The mean angle of attach $\alpha_0$ is $10^\circ$, while the angle oscillation amplitude $\alpha_s$ is $15^\circ$.
An O-grid mesh has been used in the computations, with to 512 nodes over the profile and 128 nodes in the normal direction, named medium grid in~\citet{Baldan2023}.
The element size at the leading edge is $2.0\cdot10^{-3} \,c$ while at the trailing edge is $5.0\cdot10^{-4} \,c$ according to the best practice.
The $y^+ < 1$ requirement is always satisfied for the first cell layer at the wall, for all considered operating conditions.
In Figure~\ref{img_ogrid}, a detail of the mesh is reported.
Numerical simulations are performed using ANSYS Fluent 2023R2~\citep{ANSYS}. 
The unsteady incompressible RANS equations are solved using second-order upwind discretization and second-order implicit time integration scheme.
Gradients are retrieved through a least square cell-based method, and fluxes are obtained with the Rhie-Chow momentum-based formulation. 
Pressure-velocity equations are solved using the SIMPLE method. 
A rigid motion of the entire grid is prescribed through a user-defined expression to allow the airfoil pitching, $\alpha(t) = \alpha_0 + \alpha_s \sin(\omega t)$.
The SST model with the intermittency equation~\citep{Menter2004} is employed to close the RANS equations.
Each cycle is discretized with 3\,600 time steps.
All the simulations are evolved until time convergence is reached between subsequent cycles, meaning that the obtained solution overlaps.

The design space is defined by three variables to limit the dimensionality of the problem and reduce the number of performed simulations without compromising the applicability of the proposed framework.
The input parameters of the model are the free-stream velocity $V_\infty$ that can vary between 12 and 22 m/s, the pitching amplitude angle $\alpha$ that ranges from $5^\circ$ to $15^\circ$, and the reduced frequency $k$ extends from 0.1 to 0.2.
All the other parameters are kept equal to the reference test case.
\begin{table}
	\caption{\label{tab_discrepancy} Discrepancy for Latin hypercube, Sobol', and Halton sequences of the datasets in the unitary hypercube before scaling to real values.}
	\centering
	\begin{tabular}{lccc}
		\hline
		\textbf{Dataset size} & \textbf{Latin} & \textbf{Sobol'} & \textbf{Halton} \\
		30  & $5.18\cdot 10^{-3}$ & $3.00\cdot 10^{-3}$ & $2.44\cdot 10^{-3}$ \\
		40  & $2.04\cdot 10^{-3}$ & $1.52\cdot 10^{-3}$ & $1.42\cdot 10^{-3}$ \\
		50  & $2.70\cdot 10^{-3}$ & $1.07\cdot 10^{-3}$ & $1.25\cdot 10^{-3}$ \\
		100 & $7.11\cdot 10^{-4}$ & $3.36\cdot 10^{-4}$ & $2.67\cdot 10^{-4}$ \\
		216 & $2.52\cdot 10^{-4}$ & $8.49\cdot 10^{-5}$ & $8.02\cdot 10^{-5}$ \\
		\hline
	\end{tabular}
\end{table}

\begin{figure*}
	%\centering
	\captionsetup[subfigure]{labelformat=empty}
	\subfloat[30-point Sobol]{\includegraphics[width=0.245\textwidth]{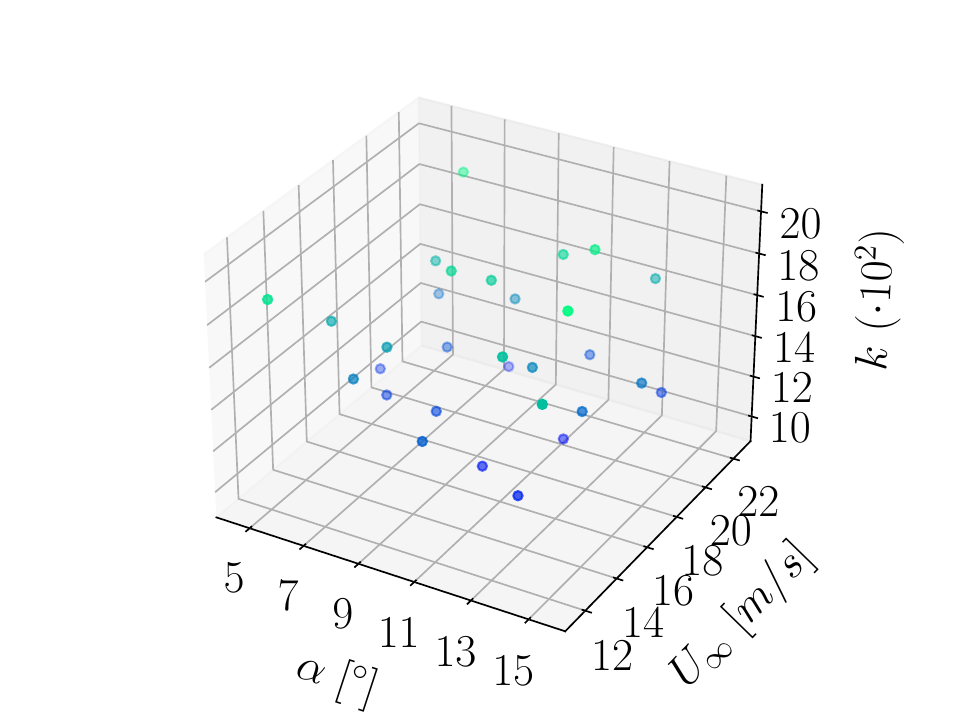}}
	\subfloat[30-point Latin Hypercube]{\includegraphics[width=0.245\textwidth]{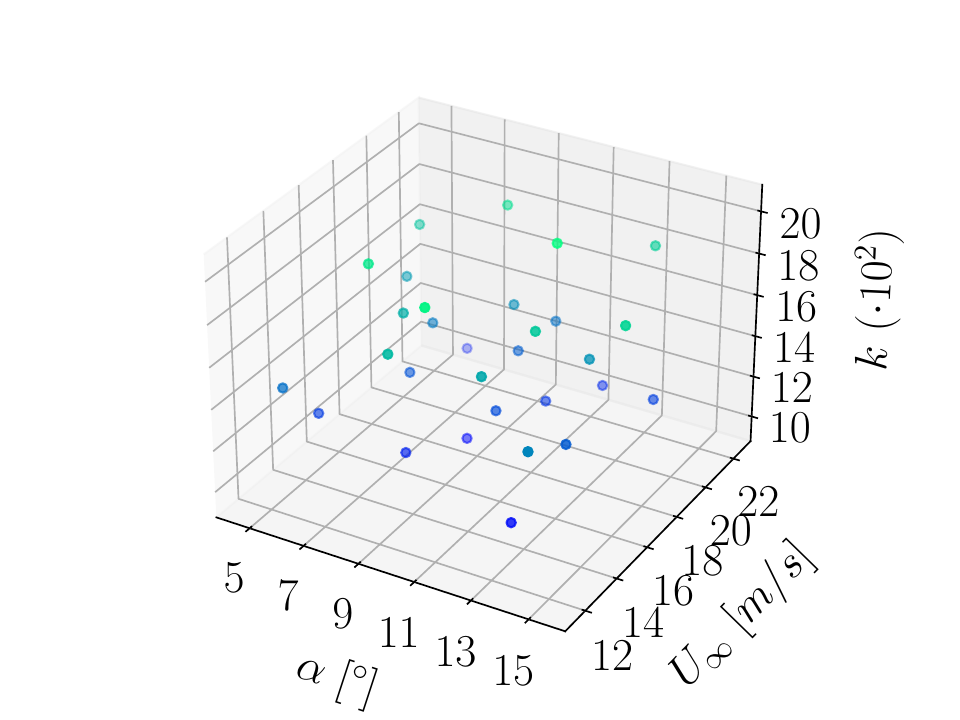}}
	\subfloat[30-point Halton]{\includegraphics[width=0.245\textwidth]{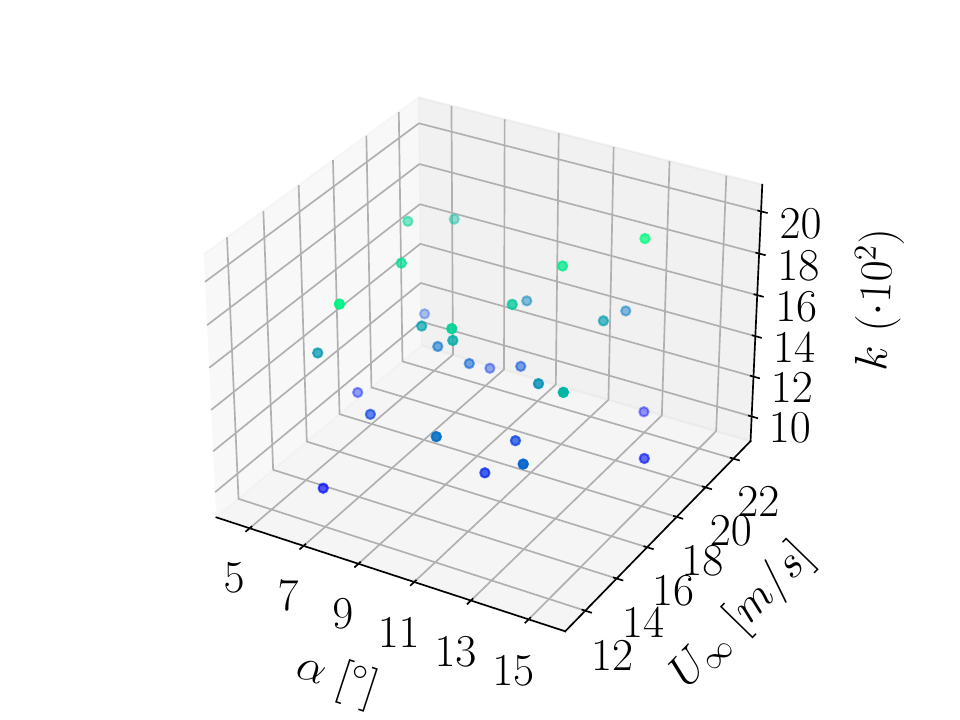}}\\
	\vspace*{5mm}
	\subfloat[40-point Sobol]{\includegraphics[width=0.245\textwidth]{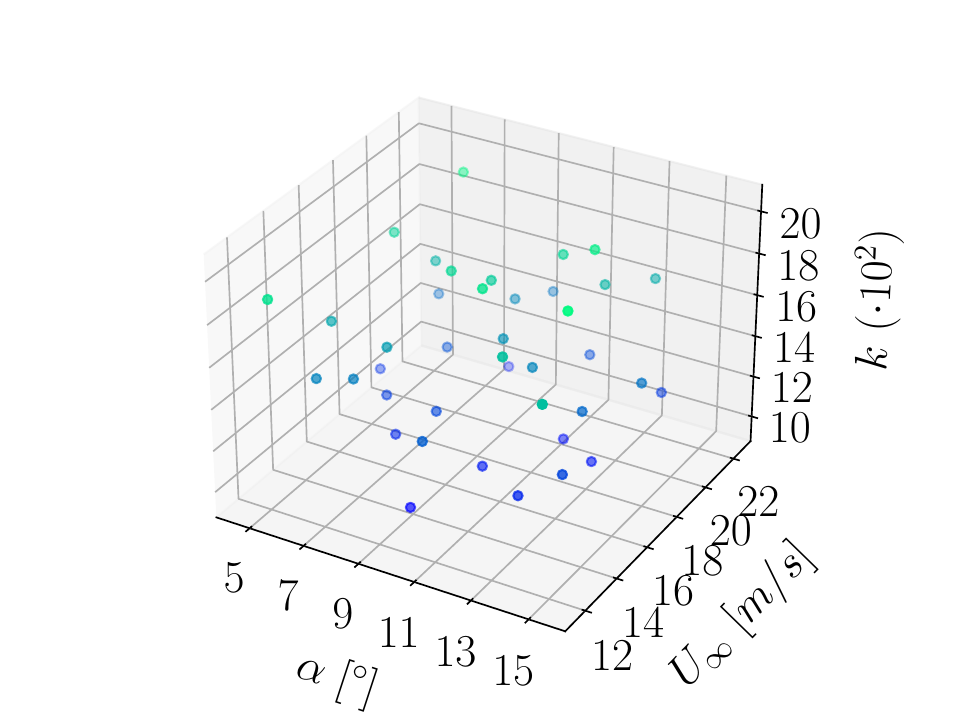}}
	\subfloat[40-point Latin Hypercube]{\includegraphics[width=0.245\textwidth]{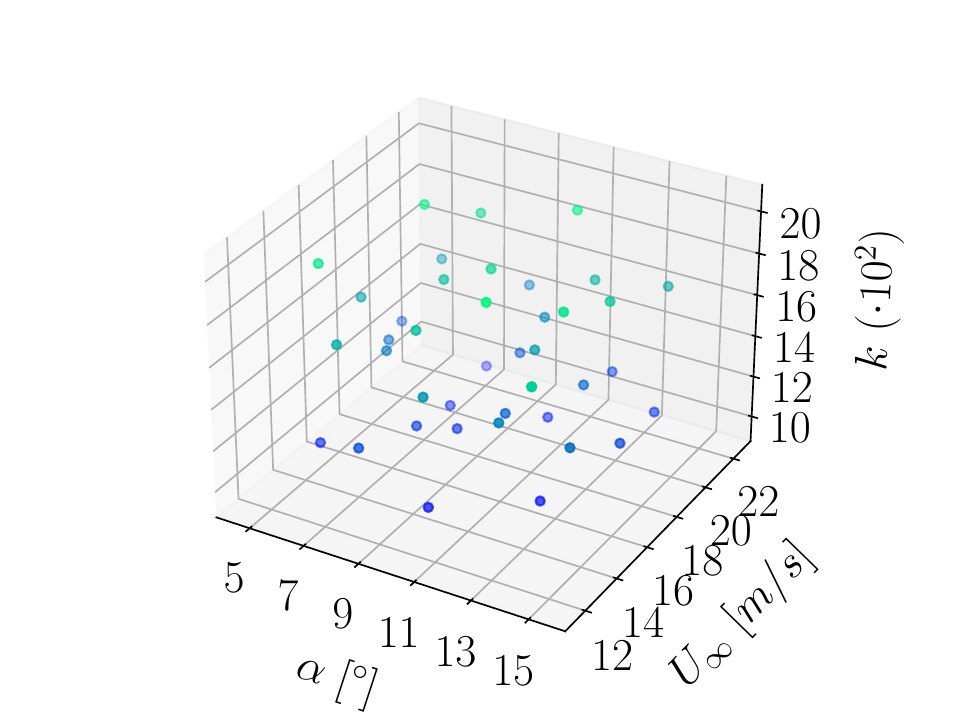}}
	\subfloat[40-point Halton]{\includegraphics[width=0.245\textwidth]{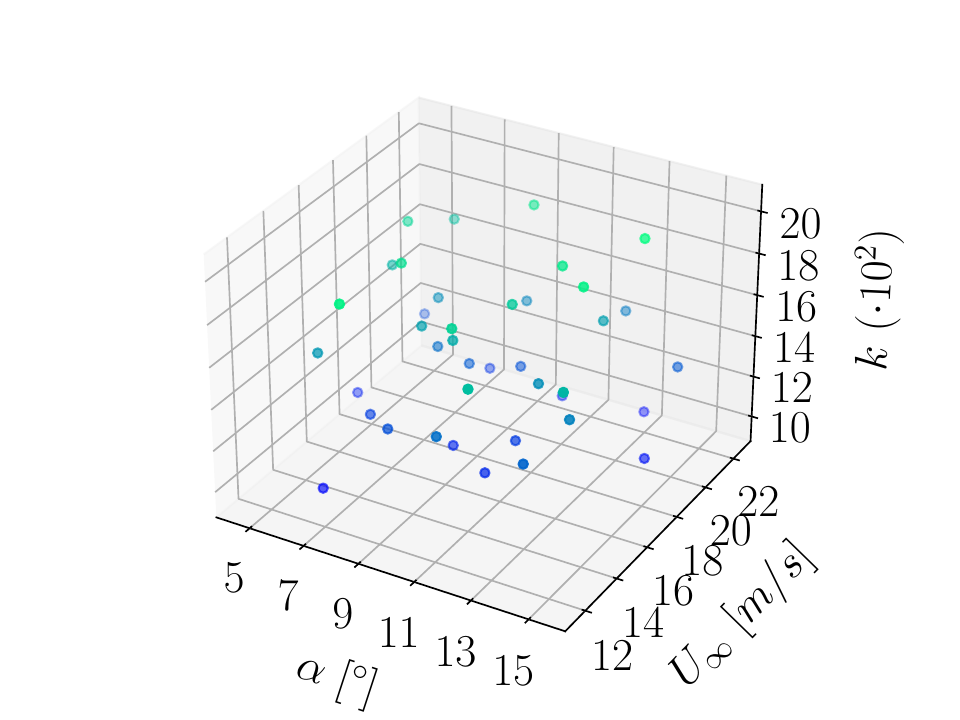}}
	\subfloat[40-point random]{\includegraphics[width=0.245\textwidth]{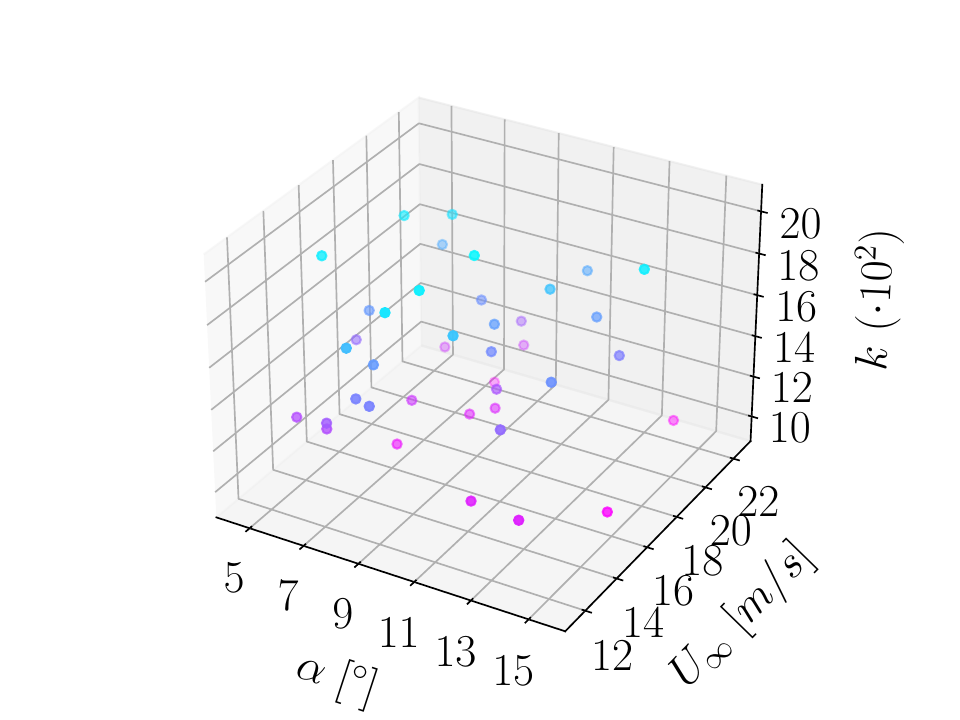}}\\                
	\vspace*{5mm}
	\subfloat[50-point Sobol]{\includegraphics[width=0.245\textwidth]{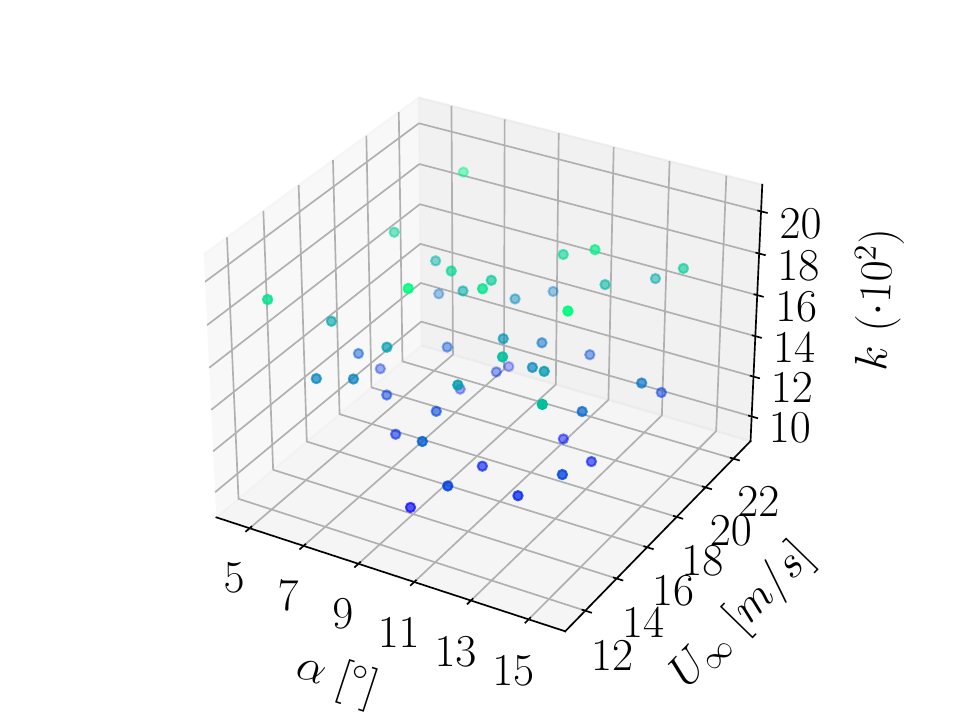}}
	\subfloat[50-point Latin Hypercube]{\includegraphics[width=0.245\textwidth]{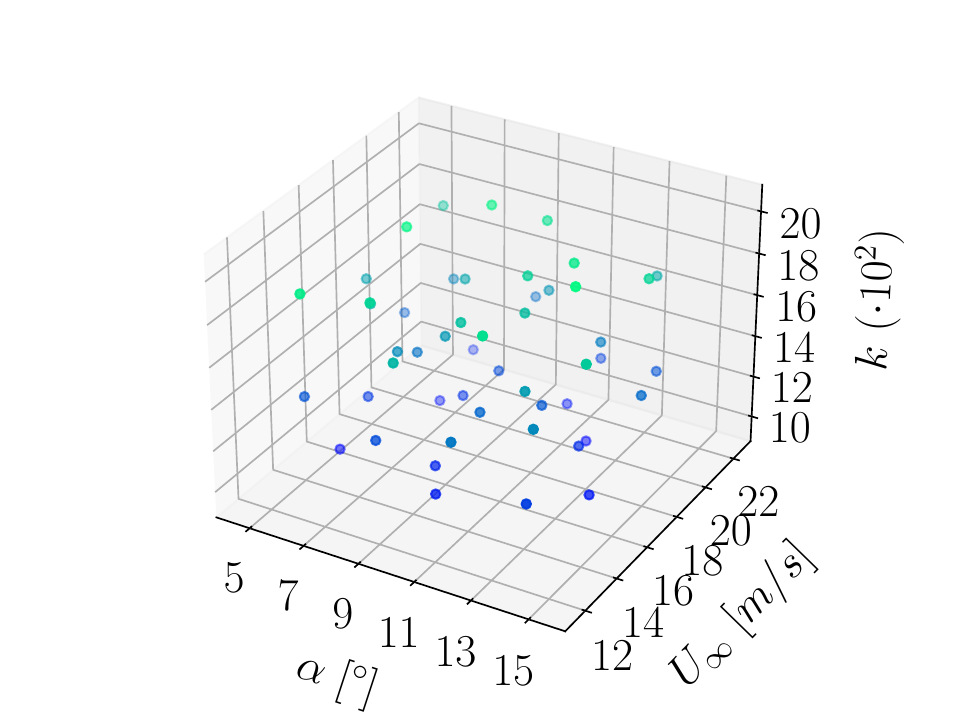}}
	\subfloat[50-point Halton]{\includegraphics[width=0.245\textwidth]{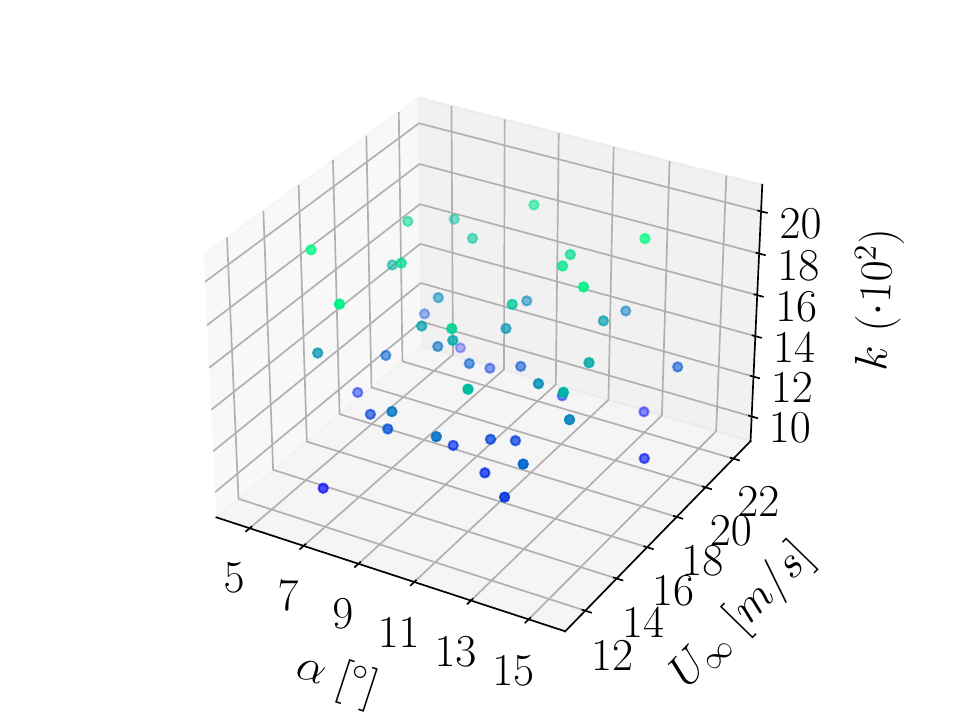}}\\
	\vspace*{5mm}
	\subfloat[100-point Sobol]{\includegraphics[width=0.245\textwidth]{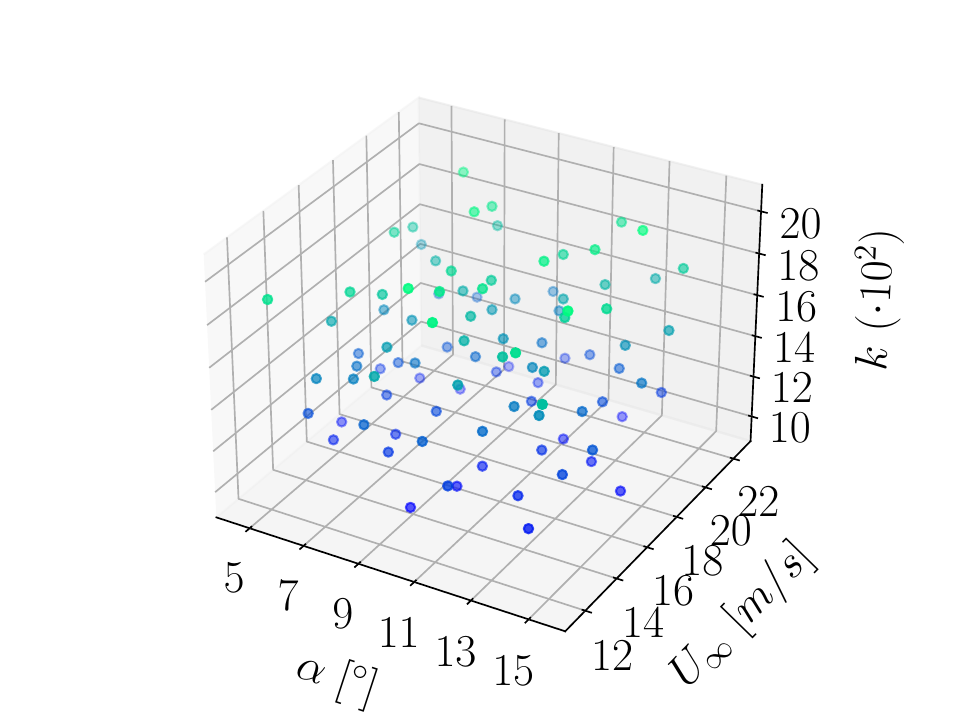}}
	\subfloat[100-point Latin Hypercube]{\includegraphics[width=0.245\textwidth]{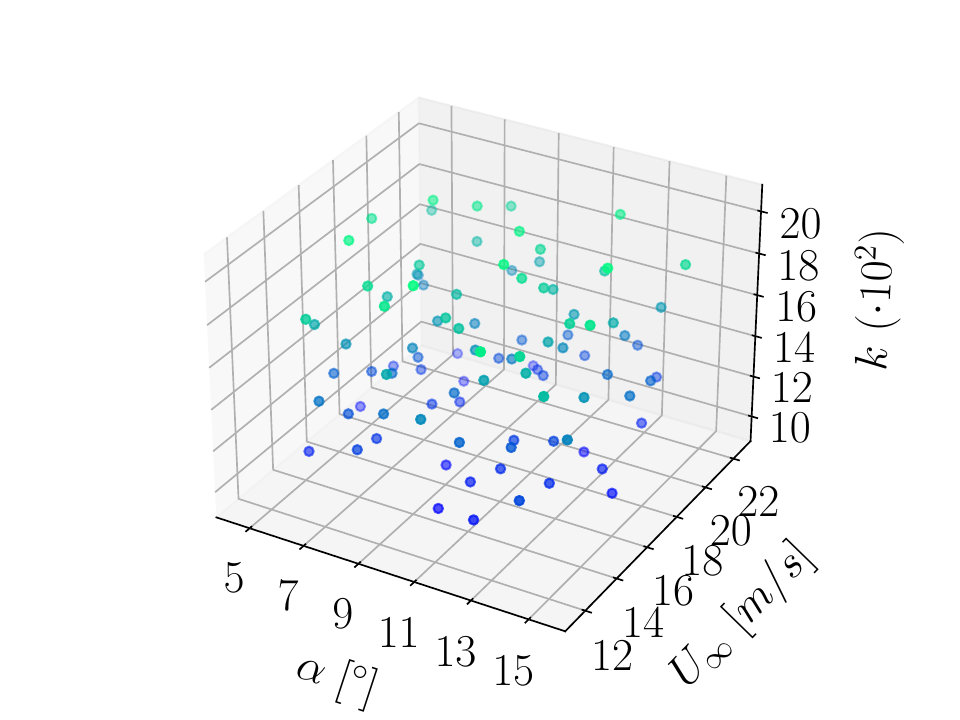}}
	\subfloat[100-point Halton]{\includegraphics[width=0.245\textwidth]{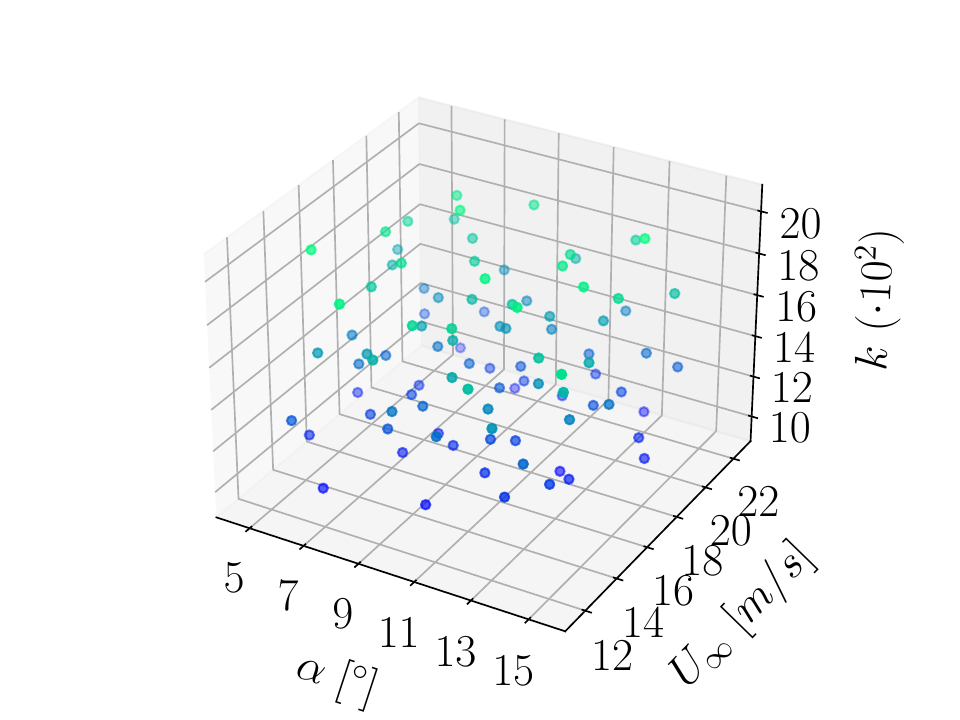}}
	\subfloat[100-point random]{\includegraphics[width=0.245\textwidth]{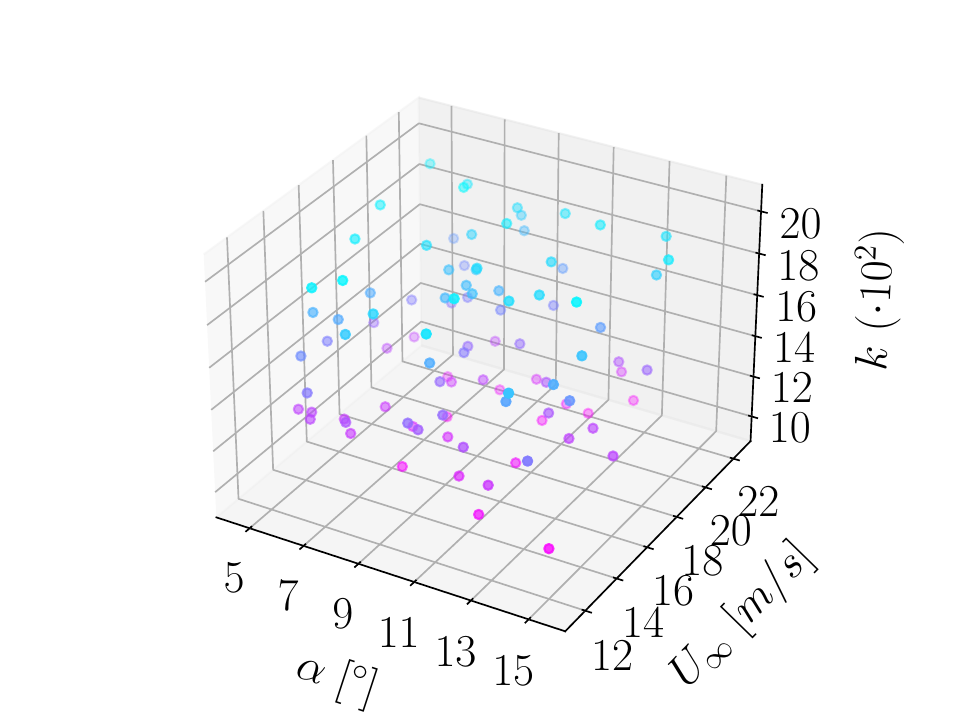}}\\
	\vspace*{5mm}
	\subfloat[216-point Sobol]{\includegraphics[width=0.245\textwidth]{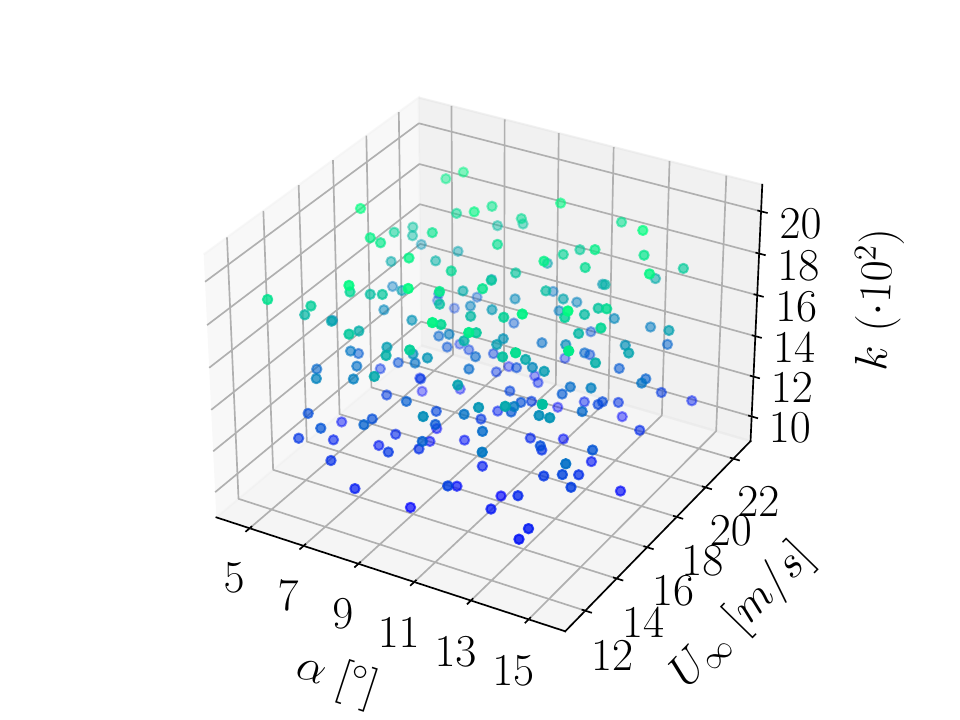}}
	\subfloat[216-point Latin Hypercube]{\includegraphics[width=0.245\textwidth]{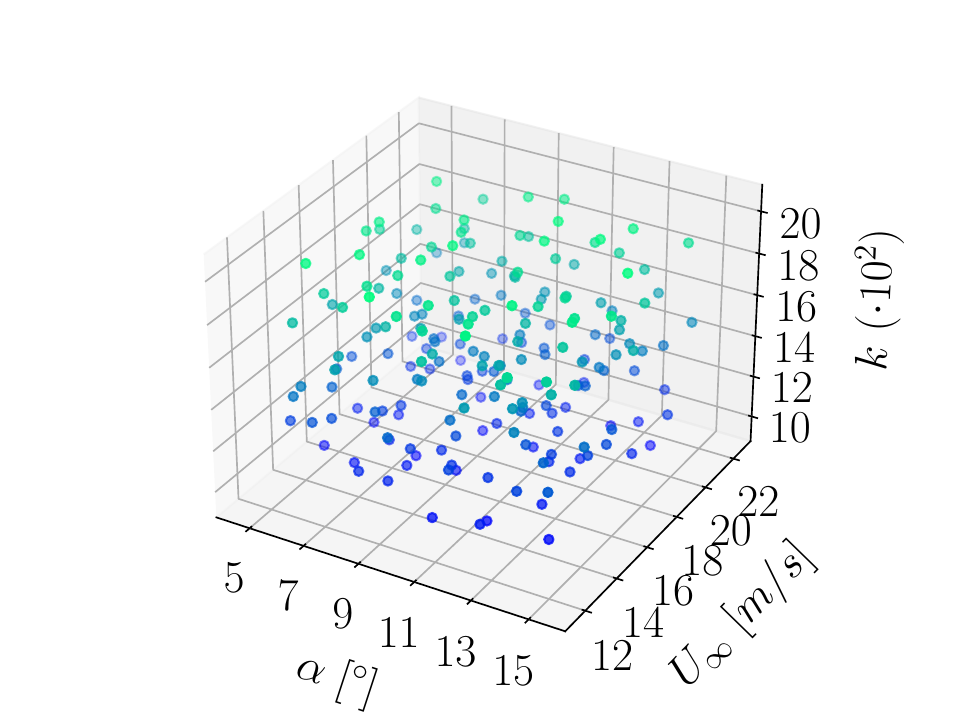}}
	\subfloat[216-point Halton]{\includegraphics[width=0.245\textwidth]{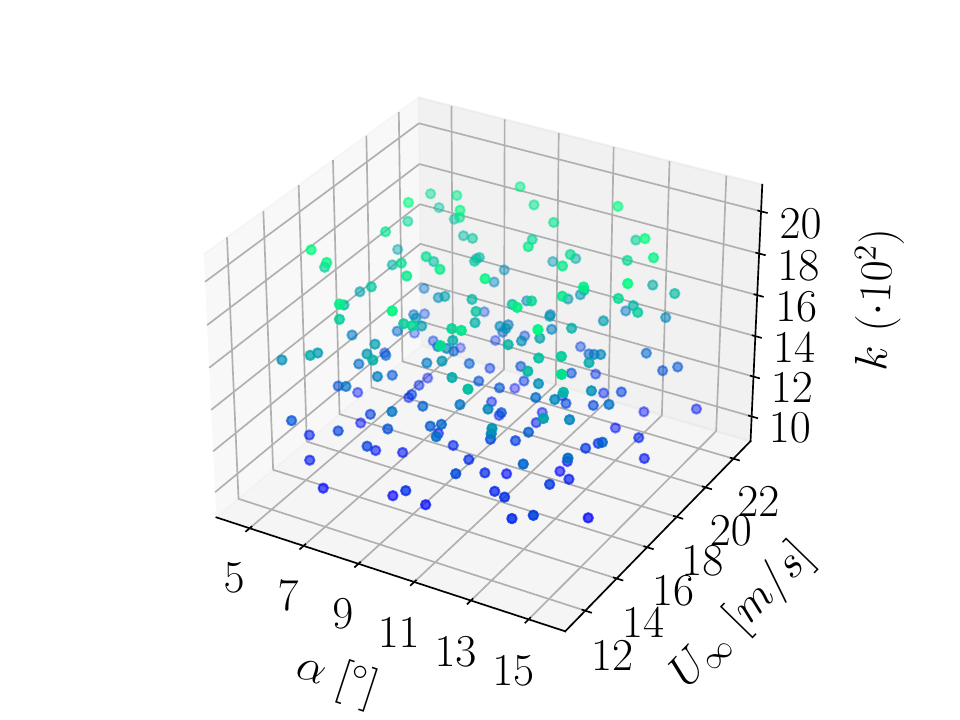}}
	\subfloat[216-point uniform grid]{\includegraphics[width=0.245\textwidth]{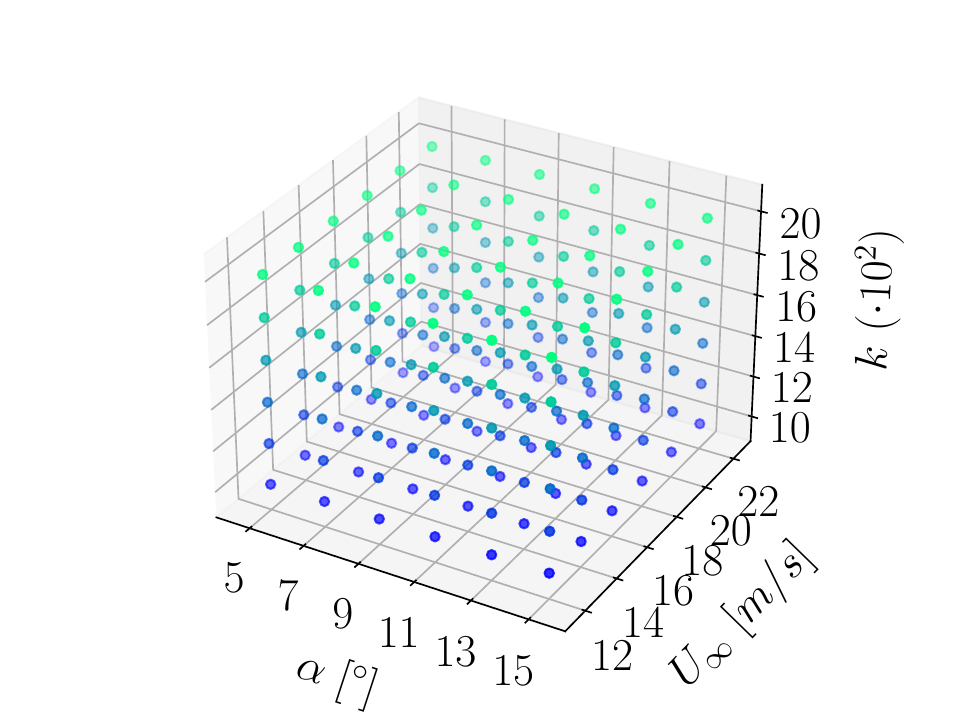}}\\
	\vspace*{5mm}
	
	\caption{Sobol, Latin Hypercube, Halton, random, and uniform grid point datasets. Training datasets are represented in blue-green gradient while test datasets in cyan-pink gradient}
	\label{img_datasets}
\end{figure*}

To generate the training distribution, a uniform grid has been used to cover the input space utilizing 6 points for each variable, namely 216 points in total.
Unfortunately, it is known that the uniform grid distribution performs poorly in terms of discrepancy especially for a limited number of points and large dimensionality~\citep{Loyola2016}.
To overcome this limitation, low-discrepancy sequences have been used.
Low-discrepancy distributions, also called quasi-random sequences, are sequences of points strategically generated to achieve a more uniform coverage of the design space compared to purely random sequences. 
The objective is to minimize discrepancies or irregularities in the distribution of points, especially in higher dimensions.
Unlike truly random sequences, low-discrepancy sequences systematically distribute points across different dimensions to create a more even sampling of the input space. 
This regular and deterministic pattern is advantageous in numerical methods and simulations where a more uniform sampling contributes to improved accuracy.
In this work, Sobol', Latin hypercube, and Halton sequences~\citep{Kocis1997} are used for training with different sizes: 30, 40, 50, 100 and 216. 
The discrepancy of each distribution is reported in Table~\ref{tab_discrepancy}.
The model validation dataset is not obtained by splitting the training dataset to preserve the low-discrepancy property of the employed sequences.
The model's performance is evaluated using two new sets of data, each containing 40 and 100 random points, respectively.
Figure~\ref{img_datasets} provides a graphical representation of all employed datasets.
Training datasets are reported in blue-green gradient while test datasets are in cyan-pink gradient.

\begin{figure}
	\centering
	\includegraphics[width=0.5\textwidth]{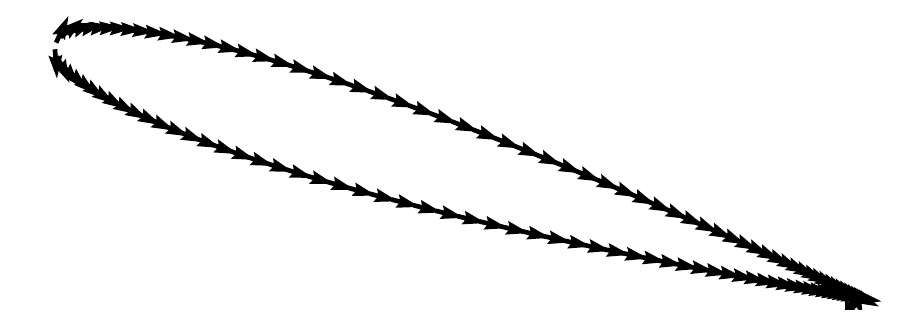}\\
	\caption{Example of the curvilinear coordinate employed in the ${C_f}$ computation. The origin is located at the leading edge while the end is posed in the middle of the trailing edge for both sides of the airfoil.}
	\label{img_geometric_direction}
\end{figure}

\begin{figure}
	\centering
	\includegraphics[width=0.5\textwidth]{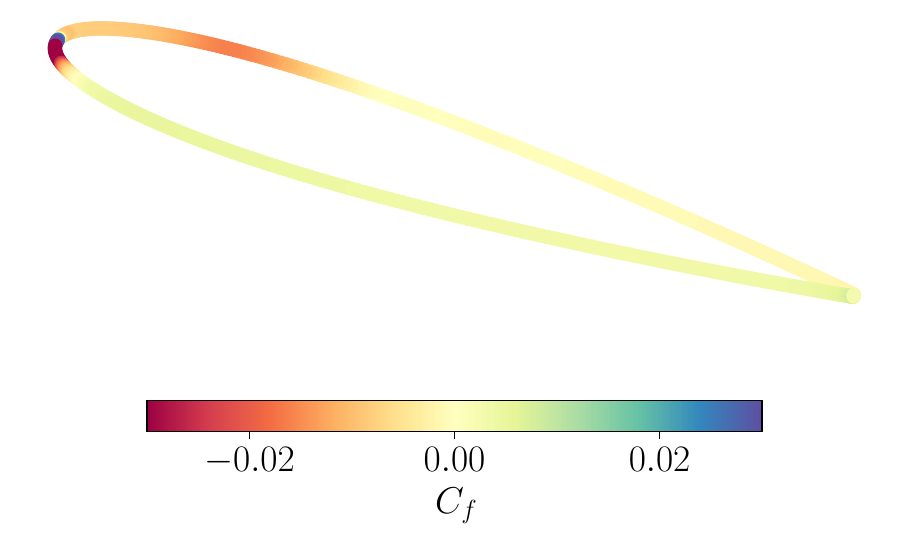}\\
	\caption{Example of the post-processed ${C_f}$ distribution. It is noticeable the discontinuity at the leading edge due to the origin of the coordinate system and the recirculation portion on the suction side of the profile due to the backward advection of the dynamic stall vortex.}
	\label{img_cf_example}
\end{figure}

\begin{figure}
	\centering
	\subfloat{\includegraphics[width=0.625\textwidth]{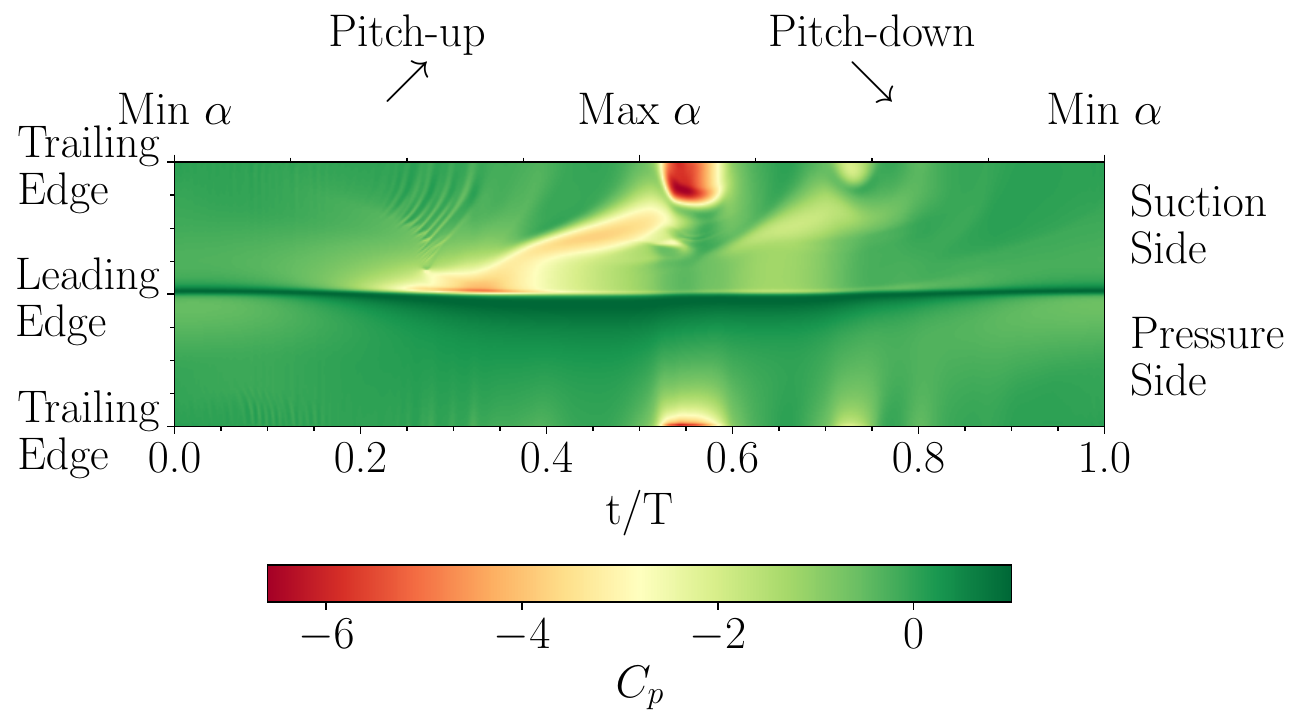}}\\
	\vspace*{5mm}
	\subfloat{\includegraphics[width=0.625\textwidth]{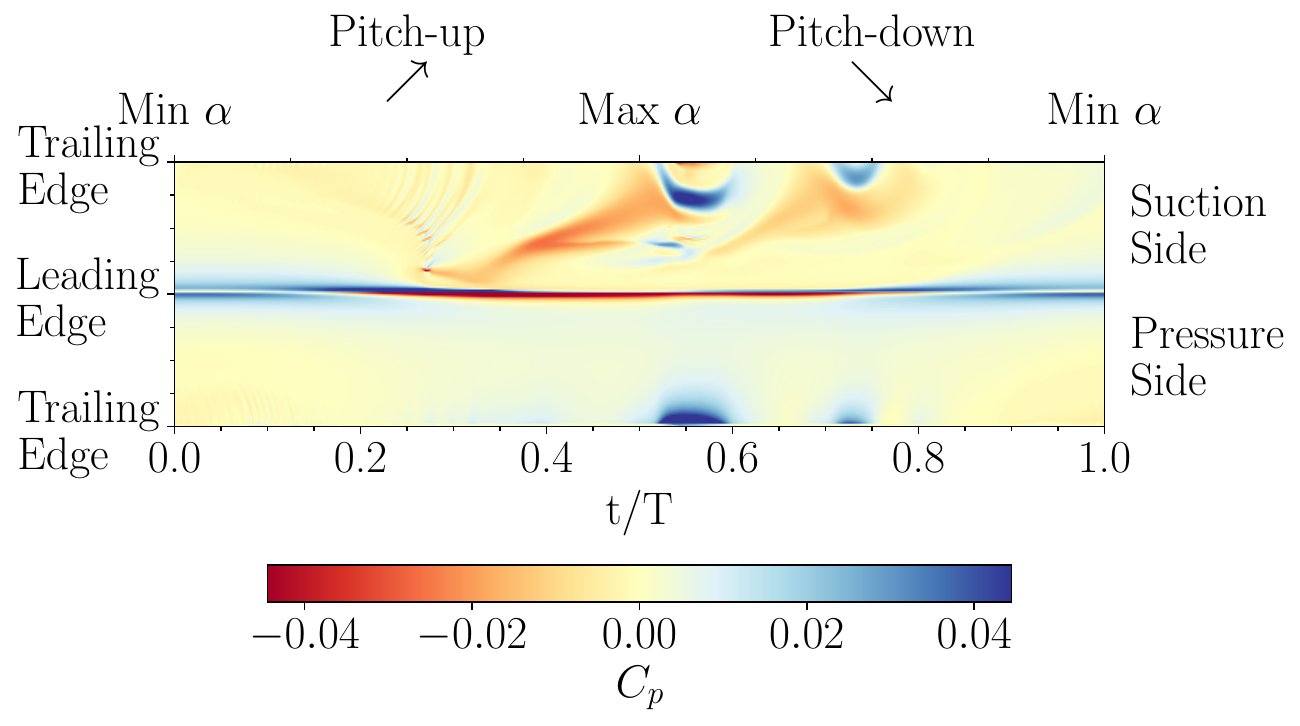}}
	\caption{Example of spatio-temporal contour of the post-processed ${C_p}$ and ${C_f}$ distributions over an entire pitching cycle.}
	\label{img_cpcf_explanation}
\end{figure}

The outputs of each simulation, which populates the datasets, are the pressure coefficient and the skin friction coefficient distributions over the profile for each time step in one pitching cycle.
The unstructured output from Fluent is mapped to the structured airfoil mesh using a k-d tree to search the nearest neighbors~\citep{Virtanen2020}.
This allows to organize the output in a 3-dimensional array and to compare results for different simulations. 
The first direction corresponds to the time step and ranges from 1 to 3\,600.
The second direction represents the airfoil surface coordinate.
Indexes from 1 to 256 refer to  data over the suction side of the profile while, from 257 to 512 to the pressure side.
The size of the last direction is 2 and stores physical data, namely $C_p$ and $C_f$.
If the full dataset is considered, an additional dimension is present that identifies the simulation index.
The post-processed value of the skin friction in each point is equal to the signed magnitude.
The sign is set according to the direction computed from the $x$ and $y$ components of the $C_f$.
The positive geometric direction is defined as the curvilinear coordinate starting from the leading edge towards the trailing edge for both sides of the airfoil.
An example is available in Figure~\ref{img_geometric_direction}.
If the geometric direction and the $C_f$ direction computed from the numerical simulations coincide, the sign is positive. Otherwise, the sign is negative.
This allows to highlight the regions where the flow is detached and helps the neural networks to identify stalled portions.
Figure~\ref{img_cf_example} highlights two relevant features: the discontinuity at the leading edge due to the origin of the coordinate system and the recirculation portion on the suction side of the profile due to the backward advection of the dynamic stall vortex.
Finally, in Figure~\ref{img_cpcf_explanation}, an example of spatio-temporal contour of the post-processed ${C_p}$ and ${C_f}$ distributions over an entire pitching cycle is available, highlighting the pitch-up and pitch-down phases and how the profile is mapped in the structured array.

\section{Neural network frameworks} \label{sec_neural_network_architectures}

\begin{figure*}
	\centering
	\subfloat[MLP Single\label{img_mlp_single}]{\includegraphics[width=0.4\textwidth]{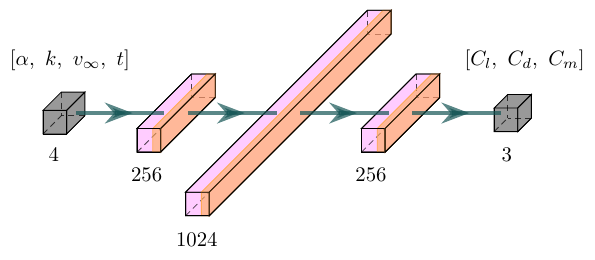}}
	\hfill
	\subfloat[MLP All\label{img_mlp_all}]{\vspace*{-2mm}\includegraphics[width=0.55\textwidth]{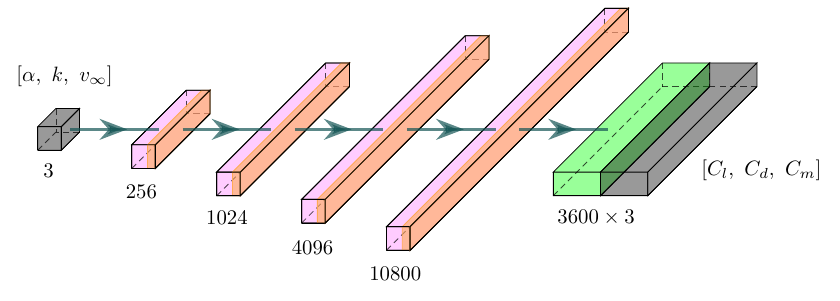}}\\
	\vspace*{10mm}
	\subfloat[CNN\label{img_cnn}]{\includegraphics[width=\textwidth]{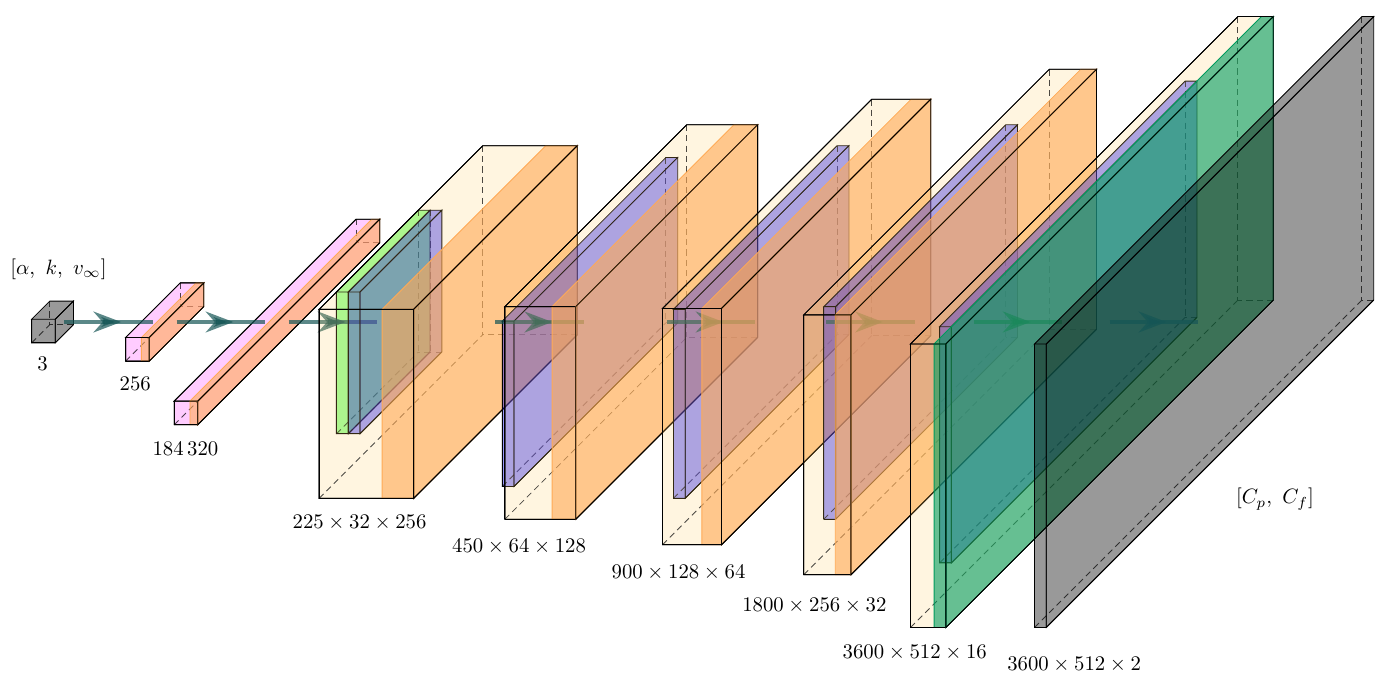}}\\
	\vspace*{5mm}
	\subfloat{\includegraphics[width=0.9\textwidth]{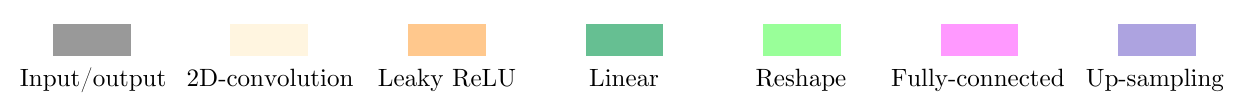}}\\
	\caption{Neural network framework representations. Input parameters are: ${\alpha}$ angle oscillation amplitude, ${k}$ reduced pitching frequency, and ${V_\infty}$ free-stream velocity. When present, ${t}$ represents the time instant over the pitching period.}
\end{figure*}

The term multilayer perceptron (MLP) is commonly used to identify feed-forward artificial neural network, consisting of fully connected neurons with a non-linear activation function, made of at least three layers~\citep{Cybenko1989}.
A convolutional neural network (CNN), instead, is a regularized type of feed-forward neural network which  optimizes the filters, or kernels, through automated learning, whereas in traditional algorithms these filters are user-defined~\citep{Albawi2017}.

The model implementation and the training pipeline rely on the TensorFlow 2.11 deep-learning framework~\citep{tensorflow}.
Three different neural network frameworks are used in this work.
The first two are MLPs. A schematic representation is reported in Figures~\ref{img_mlp_single}~and~\ref{img_mlp_all}. 
The third one is a CNN and is schematized in Figure~\ref{img_cnn}.
The level of complexity increases moving from the first to the last one.
The operating condition input for all frameworks is made of three parameters, namely $\alpha$, $V_\infty$, and $k$. All parameters are rescaled in the range $[0, 1]$ using their minimum and maximum values.
The additional input parameter time $t$ is present in the first framework.
Time is scaled using the oscillation period, meaning that $t$ can only assume values in the well-known range $[0, 1]$.

The output is different for all frameworks. 
In the first one, the values of $C_l$, $C_d$, and $C_m$ are generated at each time level. Moving to the next framework, the same coefficients are computed but for the entire cycle leading to an output array of size $3\,600\times3$. 
Finally, in the CNN the output are the pressure and skin friction distributions over the airfoil for all time steps of size $3\,600\times512\times2$. 
Starting from the output distributions, the force and moment coefficients are straightforwardly retrieved from the airfoil geometry.

The simplest implementation, corresponding to the first framework, consists in an MLP neural network that predicts $C_l$, $C_d$, and $C_m$ for a given operational condition and one temporal instant.
For this reason, in the following is referred as \textit{MLP Single}.
It is characterized by three hidden layers of size 256, 1024, and 256 respectively.
The activation function is Leaky ReLU (Rectified Linear Unit) which is commonly used to introduce non-linearity. 
The Leaky ReLU activation function allows a small, positive gradient when the input is negative, helping to address the vanishing gradient problem~\citep{Xu2015}.
Mathematically, Leaky ReLU is defined as $f(x)=\max(\gamma x,x)$, where $\gamma$ is set equal to 0.1.

Considering now the second MLP neural network, the structure is similar to the previous one since it always presents three hidden layers and adopts the same activation function. 
The differences concern the layer sizes that are: 256, 1\,024, and 4\,096. The last layer has size 10\,800 that is reshaped to $3\,600\times3$ to cope with the output size.
Due to the capability of predicting all the coefficients over one cycle is called \textit{MLP All} for brevity.

Lastly, the Convolutional Neural Network (CNN) is considered.
The first two hidden layers are formed by fully connected and Leaky ReLU-activated of size 256 and $184\,320$.
After, it is reshaped to $45\times16\times256$.
The first up-sampling layer increases the number of data points in the first direction by a factor of 5 while for the second direction they are doubled, leading to the first convolution layer with leaky ReLU activation.
Four up-sampling and convolution layers follows, where the feature dimension is halved, and the other two directions are doubled until the output size is reached.
The last convolution layer that reduces the 16 features to the pressure and friction distributions has a linear activation.
In the actual implementation, before each convolution, an additional ad-hoc implemented layer is computed to enlarge the data and ensure periodicity.
This step is very important from a physical point of view because it guarantees that there are no discontinuities over the airfoil and also that the start and the end of two cycles are coincident.
In TensorFlow, the convolution layers are applied to the extended data with padding equal to \textit{valid}.
This is the first difference with respect to the other two frameworks that are not able to grant a physical meaning to the predicted data.
Another advantage of having the $C_p$ and the $C_f$ distributions is the possibility of computing the moment coefficient with respect to an arbitrary point in the space that is not possible when directly learning the quarter-chord value.

The loss function $\mathcal{L}$ is the same in all the three frameworks and it is the mean squared error (MSE).
\begin{equation}
	\text{MSE} = \frac{1}{N} \sum_{i=1}^{N} (\hat{y}_i - y_i)^2,
\end{equation}
where $N$ is the number of data points, $\hat{y}_i$ is the predicted value for the $i$-th data point, and $y_i$ is the ground truth value for the $i$-th data point.

\begin{figure}
	\centering
	\includegraphics[width=0.5\textwidth]{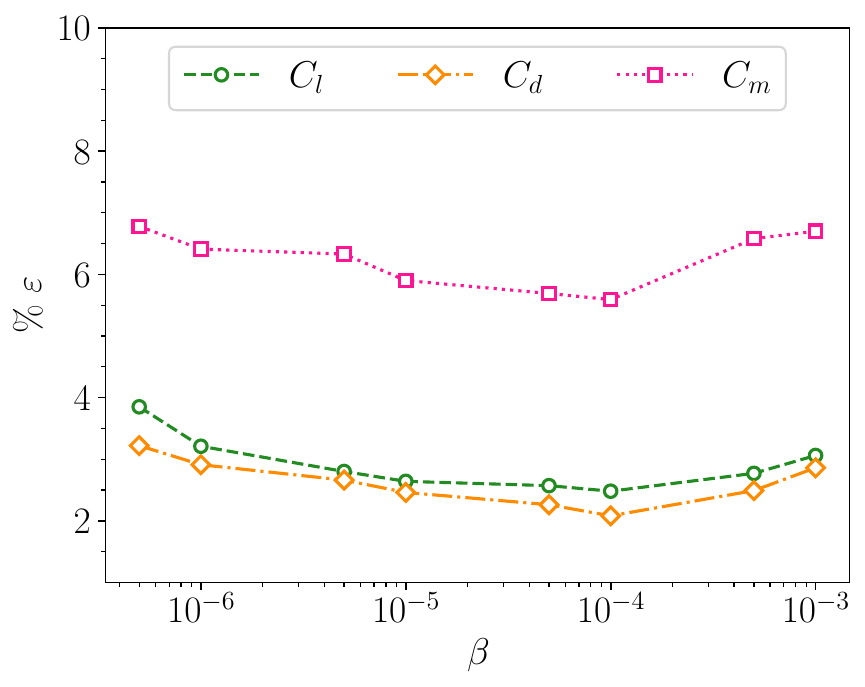}\\
	
	\caption{Influence of $\beta$ hyperparameter on the mean percentage error for $C_l$, $C_d$, and $C_m$. Training dataset is 216-point Halton and test dataset is 100-point Random.}
	\label{img_beta_hyperparameter}
\end{figure}

For the CNN, an additional loss function has been implemented and tested trying to incorporate more physical information in the fourth model.
Indeed, to build a robust model, only the data-driven approach is usually not sufficient and physics information have to be included in the model~\cite{Ribeiro2023}.
The objective is to include the mean squared force and moment coefficient error, $\text{MSE}_{\text{C}_\text{l},\text{C}_\text{d},\text{C}_\text{m}}$, in the loss function $\mathcal{L}$ in addition to the MSE presented before $\text{MSE}_{\text{C}_\text{p},\text{C}_\text{f}}$.
\begin{equation}
	\mathcal{L} = \text{MSE}_{\text{C}_\text{p},\text{C}_\text{f}} + \beta \cdot \text{MSE}_{\text{C}_\text{l},\text{C}_\text{d},\text{C}_\text{m}}
\end{equation}
The hyperparameter $\beta$ controls the influence of the coefficient error in the total loss. 
The optimal value for $\beta$ was found to be equal to $1\cdot10^{-4}$ in order to have a comparable error from the two contributions to the total loss function.
A complete analysis on the influence of $\beta$ is reported in Figure~\ref{img_beta_hyperparameter}.
The shape of the airfoil and the mesh point positions are given as input to the neural network to compute the aerodynamic loads.
To speed up computations, the \texttt{tf.function} python decorator is used.
The two models utilizing the CNN framework are distinguished based on their loss functions. The model employing the Mean Squared Error loss is referred as \textit{CNN MSE}, while the one incorporating the additional physics-based loss function is designated as \textit{CNN Coefficients}.

The Adam algorithm~\citep{Kingma2014} is employed, featuring a variable learning rate that starts at $10^{-4}$ and progressively decreases through a piecewise constant decay to $10^{-8}$. 
The training duration spans a maximum of 2\,000 epochs. 
The complete training datasets are employed, while validation is performed on the 40-point random dataset.
Training and inference wall-time for the analyzed frameworks is summarized in Table~\ref{tab_nn_time}. 
The performances are evaluated using the 216-point Halton dataset for training and the 100-point random dataset for inference. 
A single NVIDIA A100 SXM4 40GB GPU~\citep{Choquette2021} is used in all the computations.
Note that the improved loss function slightly increases the training time but does not influence the inference time still allowing a fast prediction.

\begin{table}
	\caption{\label{tab_nn_time} Training and inference wall-time for the analyzed framework. Performances are evaluated with the 216-point Halton dataset for training and the 100-point random dataset for inference. A single NVIDIA A100 SXM4 40GB GPU is used in all the computations.}
	\centering
	\begin{tabular}{lcc}
		\hline
		\textbf{Framework} & \textbf{Training time} ${[h]}$& \textbf{Inference time} ${[s]}$ \\
		MLP single & 0.24 & $2.45\cdot 10^{-3}$ \\
		MLP all & 0.32 & $3.37\cdot 10^{-3}$ \\
		CNN MSE & 3.82 & 1.10 \\
		CNN Coefficients & 4.01 & 1.10 \\
		\hline
	\end{tabular}
\end{table}

\section{Results} \label{sec_results}

\begin{figure*}
	\centering
	\includegraphics[width=0.9\textwidth]{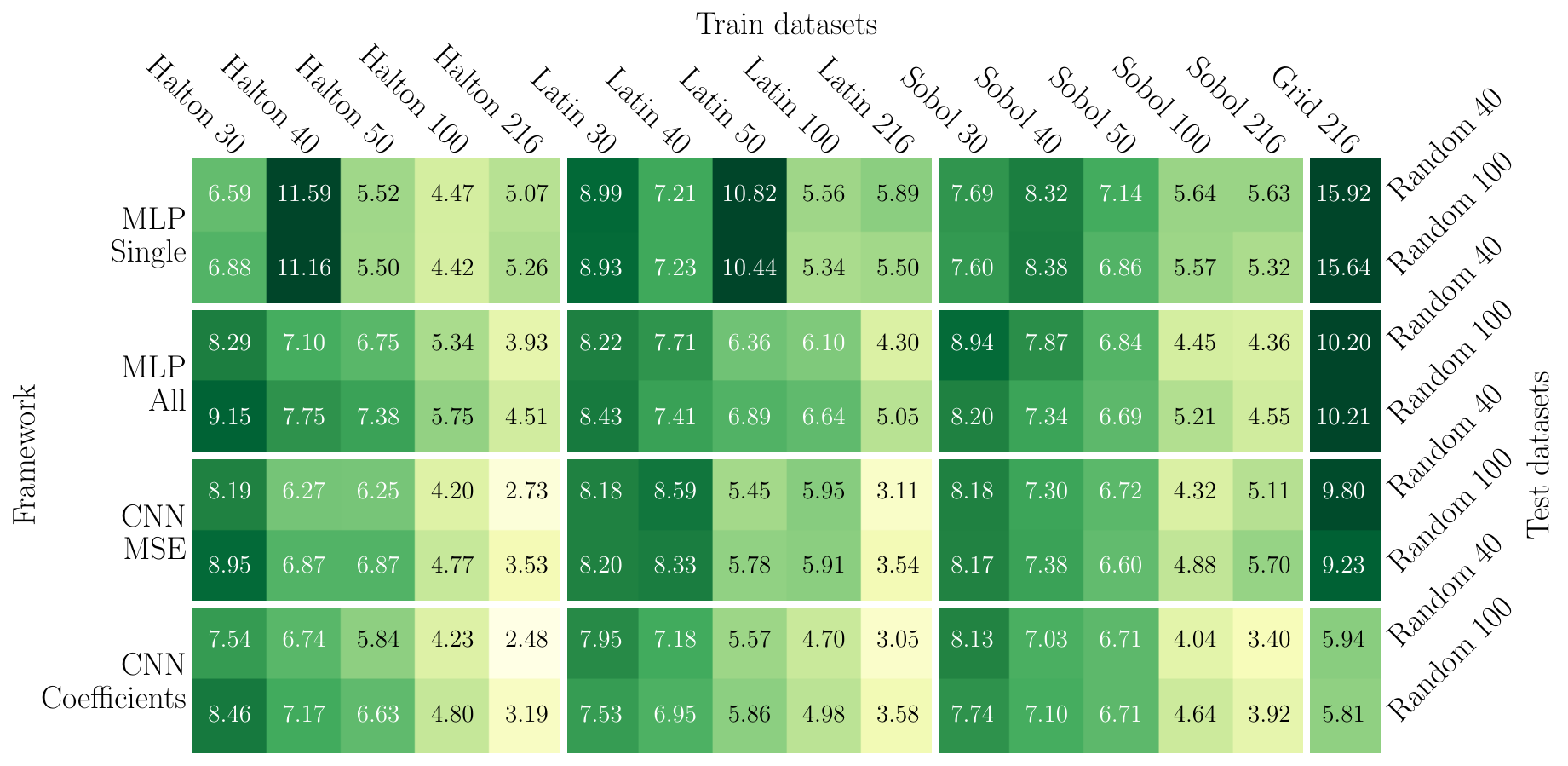}\\
	\vspace*{-.7cm}
	\includegraphics[width=0.9\textwidth]{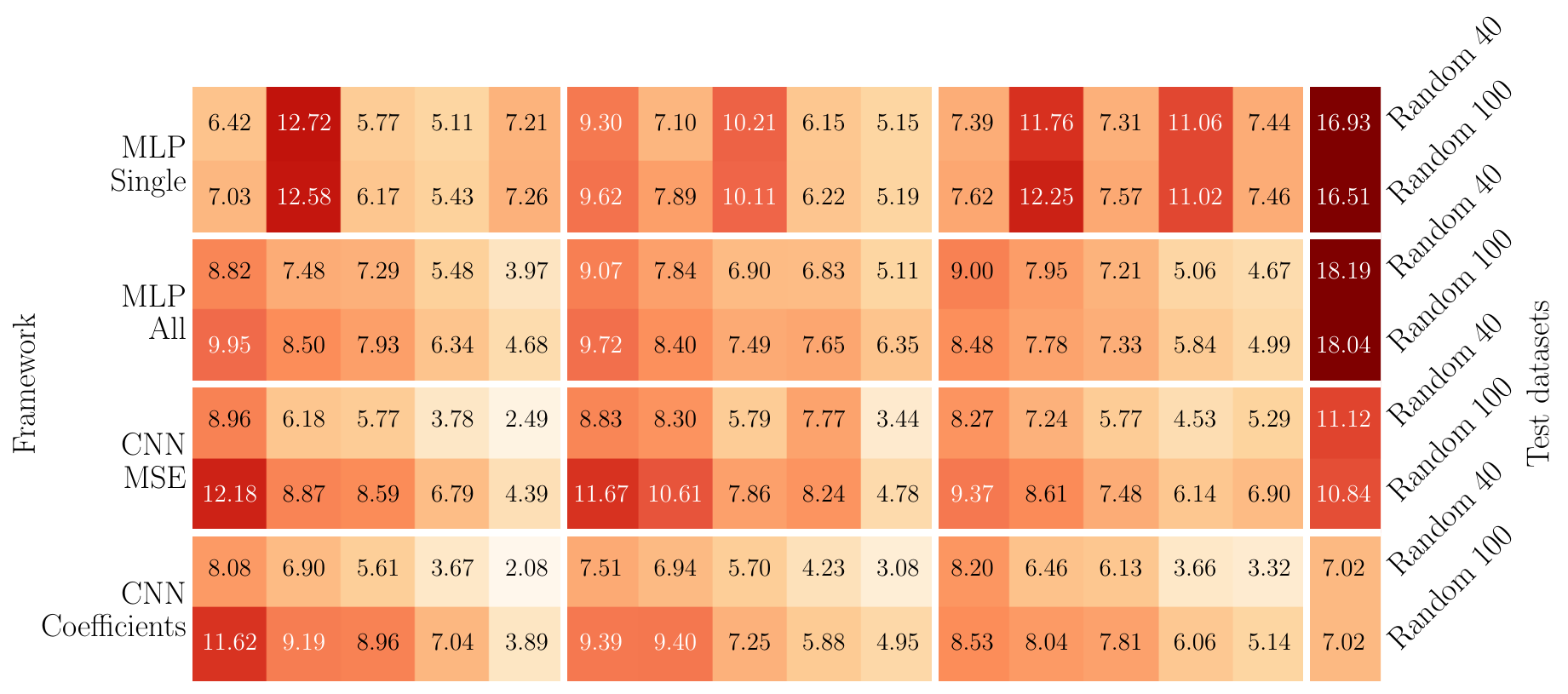}\\
	\vspace*{-.7cm}
	\includegraphics[width=0.9\textwidth]{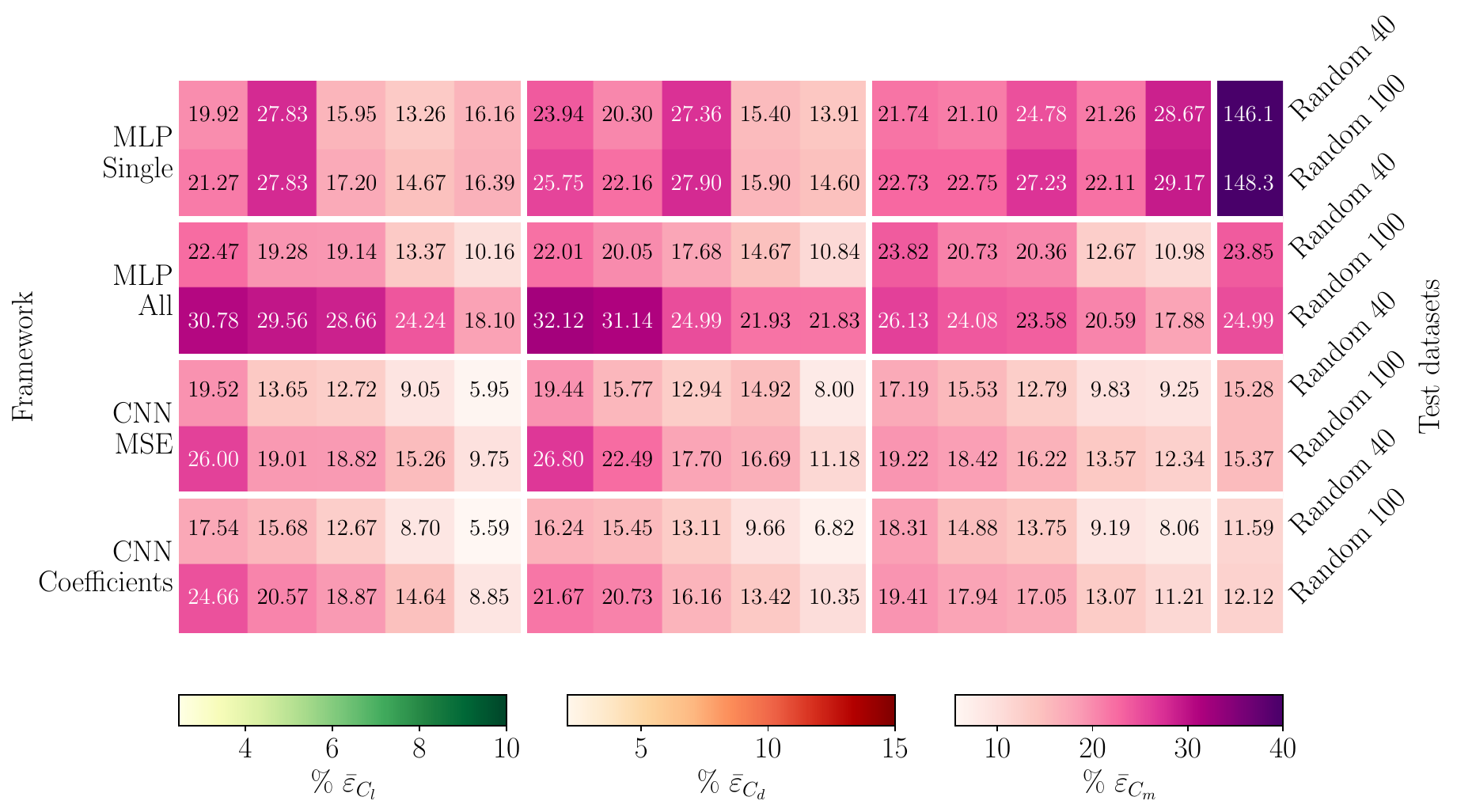}\\
	
	\caption{Mean lift, drag, and moment coefficient error (Equations~\ref{eq_mean_error_total},~\ref{eq_mean_error_snapshot}) comparison for the analyzed NN frameworks using different datasets for training and testing.}
	\label{img_mse_error}
\end{figure*}

\begin{figure*}
	\centering
	\subfloat{\includegraphics[width=\textwidth]{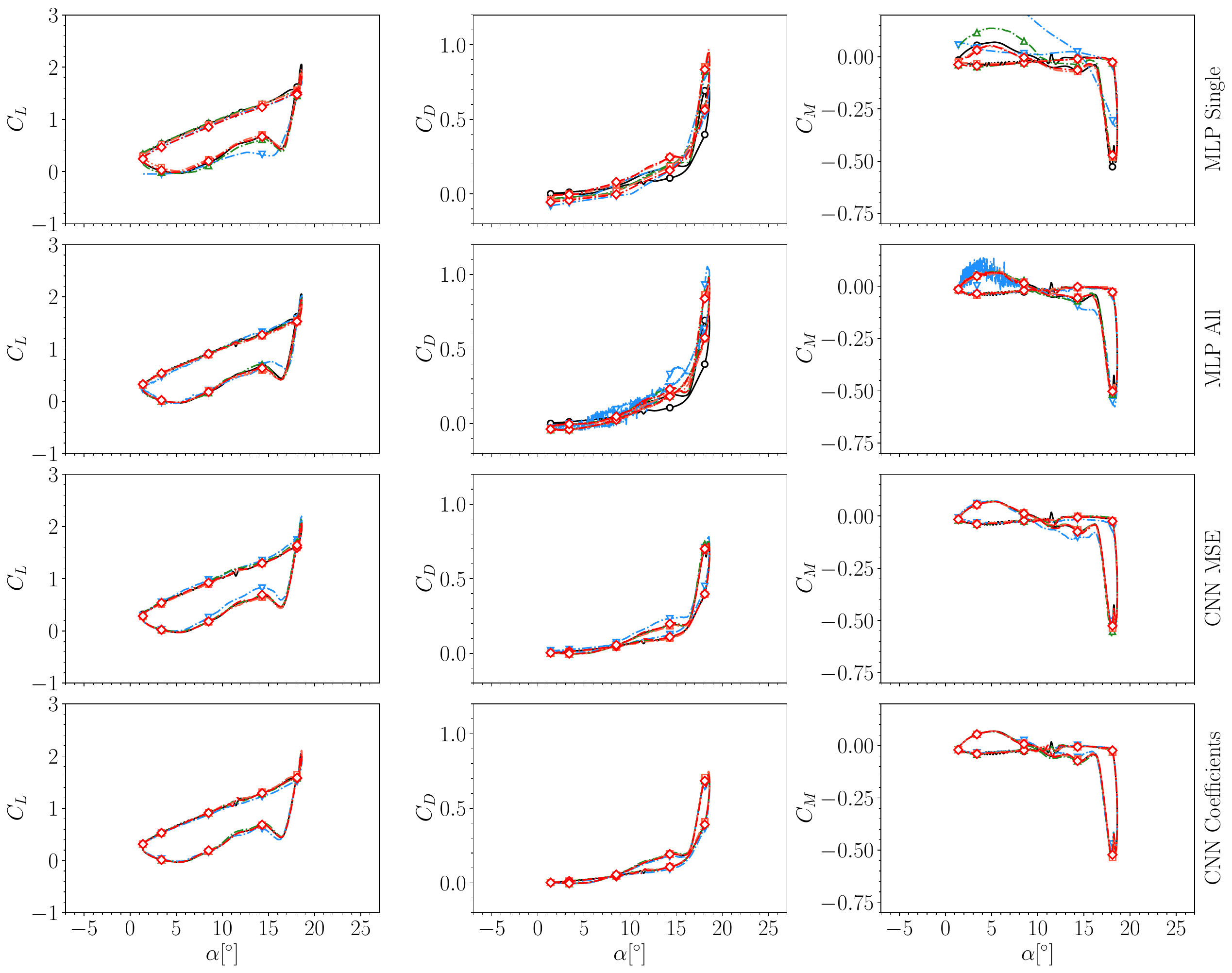}}\\
	\vspace*{5mm}
	\subfloat{\includegraphics[width=0.9\textwidth]{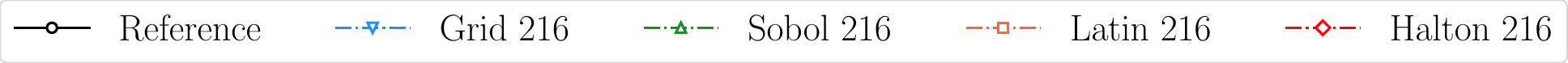}}\\
	
	\caption{Influence of quasi-random training sequence type on NN framework performances for a light dynamic stall example, corresponding to an oscillation amplitude of ${8.62^\circ}$, a 18.96 m/s free-stream velocity and a reduced frequency of 0.157.}
	\label{img_nn_coefficients_light}
\end{figure*}

\begin{figure*}
	\centering
	\subfloat{\includegraphics[width=\textwidth]{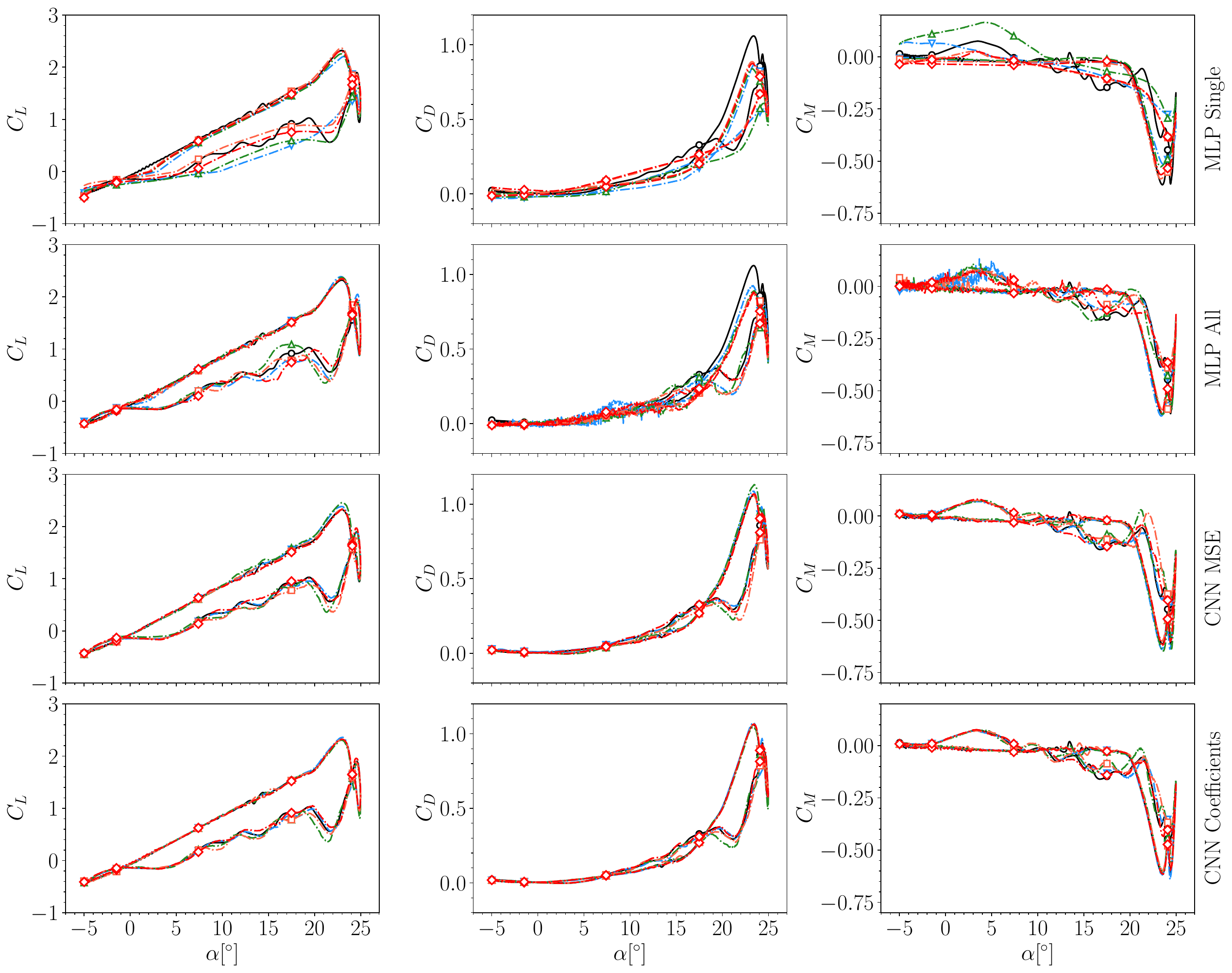}}\\
	\vspace*{5mm}
	\subfloat{\includegraphics[width=0.9\textwidth]{images/nn_coefficients_legend.pdf}}\\
	
	\caption{Influence of quasi-random training sequence type on NN framework performance for a deep dynamic stall example, corresponding to an oscillation amplitude of ${14.99^\circ}$, a 15.71 m/s free-stream velocity and a reduced frequency of 0.105.}
	\label{img_nn_coefficients_deep}
\end{figure*}

\begin{figure*}
	\centering
	\subfloat{\includegraphics[width=\textwidth]{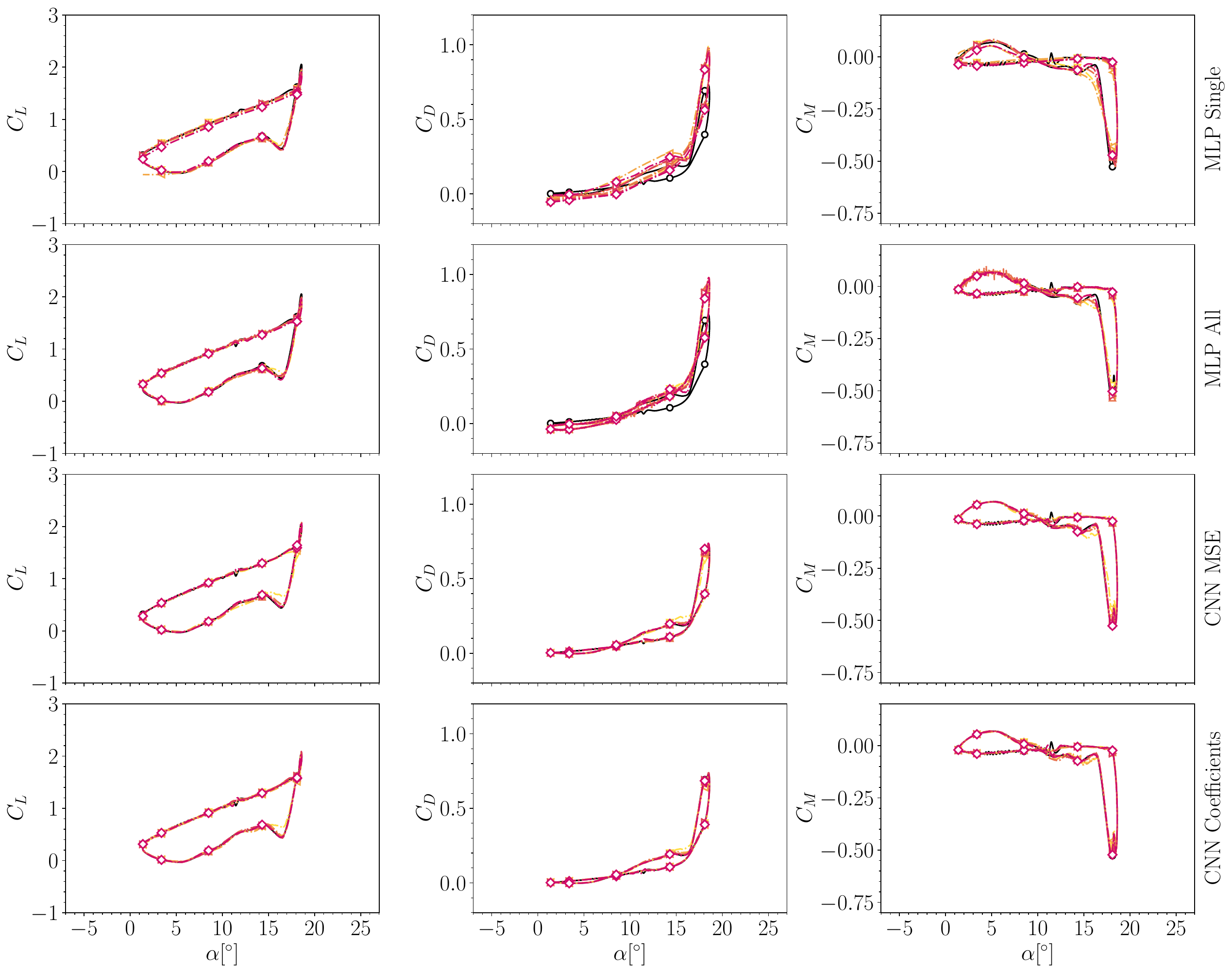}}\\
	\vspace*{5mm}
	\subfloat{\includegraphics[width=\textwidth]{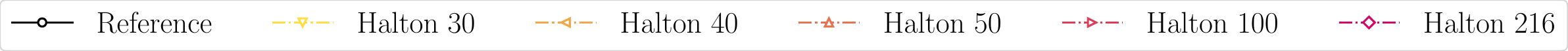}}\\
	
	\caption{Influence of Halton training sequence size on NN framework performances for a light dynamic stall example, corresponding to an oscillation amplitude of ${8.62^\circ}$, a 18.96 m/s free-stream velocity and a reduced frequency of 0.157.}
	\label{img_nn_coefficients_size_light}
\end{figure*}

\begin{figure*}
	\centering
	\subfloat{\includegraphics[width=\textwidth]{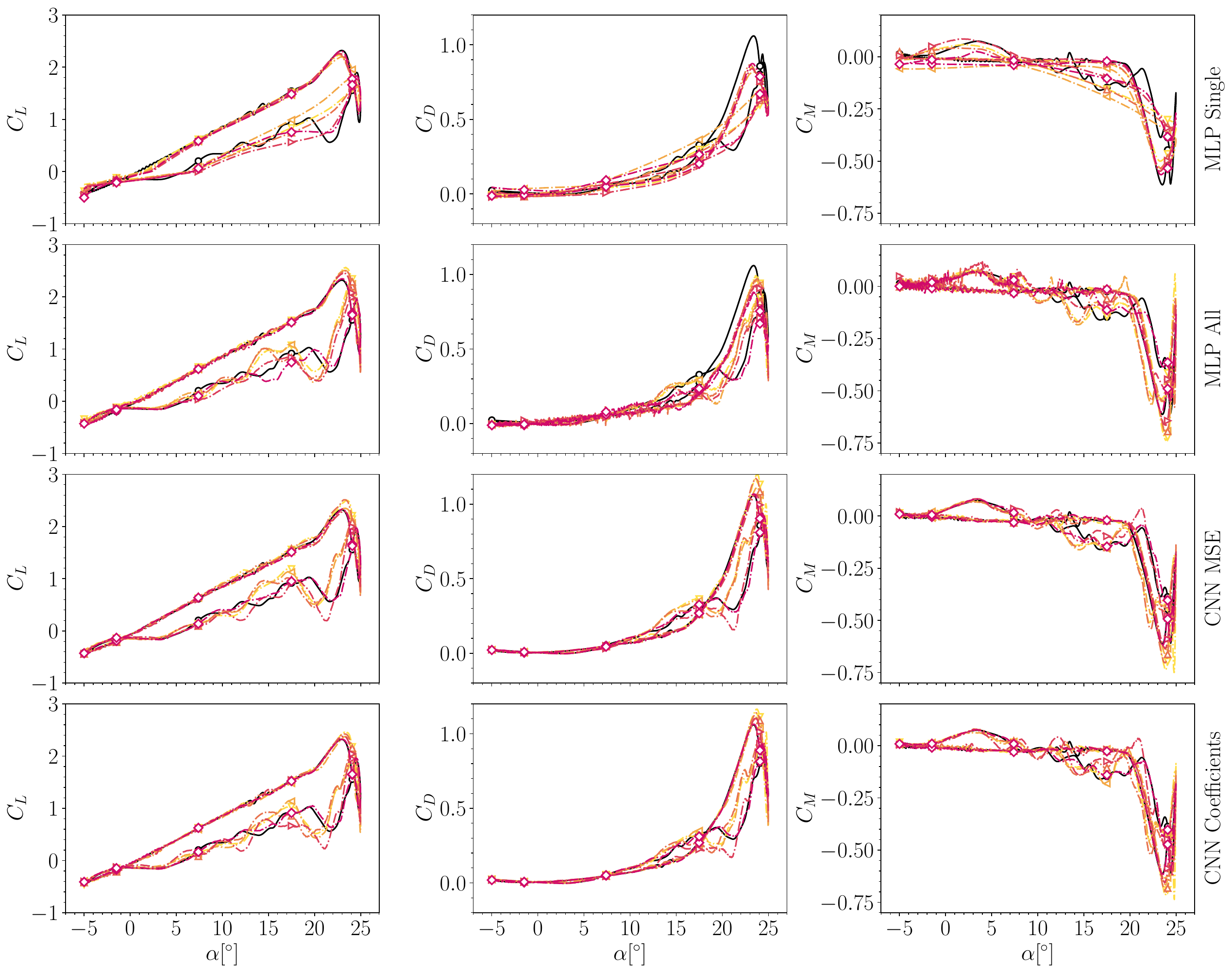}}\\
	\vspace*{5mm}
	\subfloat{\includegraphics[width=\textwidth]{images/nn_coefficients_size_legend.pdf}}\\
	
	\caption{Influence of Halton training sequence size on NN framework performances for a deep dynamic stall example, corresponding to an oscillation amplitude of ${14.99^\circ}$, a 15.71 m/s free-stream velocity and a reduced frequency of 0.105.}
	\label{img_nn_coefficients_size_deep}
\end{figure*}

This section delves into the performance of the presented neural network architectures in predicting the loads experienced by airfoils undergoing dynamic stall during pitching motion. It highlights both the strengths and the weaknesses of these networks comparing the results with high-fidelity numerical simulations. The analysis focuses on how the networks can handle unseen data and maintain accuracy across different operating conditions.

The prediction error of an entire dataset is computed as the average error of the predicted coefficients per each simulation as reported in Equation~\ref{eq_mean_error_total}.
In detail, $n_s$ is the number of simulations in the dataset and ${\varepsilon_s}^2$ is the squared error of the single snapshot per each coefficient. 
The error of the single simulation is derived as in Equation~\ref{eq_mean_error_snapshot} in which $n_t$ is the number of time steps, namely 3\,600, $\hat x_{s,t}$ is the predicted coefficient at time $t$ in simulation $s$, and $x_{s,t}$ is the reference value.
The reason for summing up the squared errors and dividing them by the sum of the reference values is linked to the occurrence of values close to zero.
Indeed, when the ground truth value is close to zero, such as the $C_m$, a small discrepancy in the predicted value can lead to large relative errors giving a wrong estimation.
\begin{align} 
	\label{eq_mean_error_total}
	\bar \varepsilon &= \sqrt{\frac{ \sum_{s=1}^{n_s} {\varepsilon_\text{s}}^2 }{n_s}}\\
	\label{eq_mean_error_snapshot}
	{\varepsilon_\text{s}}^2 &= \frac{ \sum_{t=1}^{n_t} \left(\hat x_{s,t} - x_{s,t} \right)^2 }  { \sum_{t=1}^{n_t} {x_{s,t}}^2 }
\end{align}
The mean error provides overall performance of each framework.
Since we are also interested in the performance for specific operating conditions, two test cases have been selected from the 100-point random dataset to investigate a light and a deep dynamic stall conditions.
The light dynamic stall is characterized by an oscillation amplitude of $8.62^\circ$, a 18.96 m/s free-stream velocity and a reduced frequency of 0.157.
Deep dynamic stall, instead, occurs for an oscillation amplitude of $14.99^\circ$, a 15.71 m/s free-stream velocity and a reduced frequency of 0.105.
In Figure~\ref{img_mse_error}, the mean lift, drag, and moment coefficient relative errors are compared for all the analyzed frameworks with all the training and test datasets.
The following two figures, \ref{img_nn_coefficients_light}~and~\ref{img_nn_coefficients_deep}, show the influence of the point distribution of the training datasets maintaining the number of points constant for the aerodynamic loads in the selected test cases.
Figures~\ref{img_nn_coefficients_size_light}~and~\ref{img_nn_coefficients_size_deep}, instead, investigate the effect of the point number for the Halton sequence for the same test cases.

Focusing on the \textit{MLP Single} neural network, the first aspect to notice is the sensitivity to the point distribution.
Indeed, on the uniform grid the $C_m$ error is not acceptable and it is several times larger than the other datasets.
Moreover, for the same low discrepancy sequence, increasing the number of training points does not results in improving the accuracy, since the error is not following any particular trend.
Considering now the predicted coefficients for the chosen test cases we can see that the trends are well captured for lift and drag.
The pitching moment, instead, is in good agreement with the reference data only with the Halton and Latin hypercube datasets, even if for the deep dynamic stall case is completely missing the oscillations in both upstroke and downstroke phases. 
The framework suffers particularly in the $C_m$ prediction.
One of the reasons leading to large mean error is that the predicted curve is not a closed loop suggesting that the underlying physics is not learned.

The \textit{MLP All} neural network outputs the force ad moment coefficients over an entire cycle.
According to the mean errors reported in Figure~\ref{img_mse_error}, the model better catches the physics of the problem.
In particular, it is less sensitive to the point distribution used for the training and also the uniform grid distribution shows errors comparable to low discrepancy ones.
Furthermore, enlarging the training datasets results in monotonically decreasing errors as expected.
Despite the promising properties shown in the mean error map, when model predictions are analyzed, an additional issue is identified.
Especially the drag and moment curves show spurious non-physical oscillations at low angle of attack making the model not suitable for the applications.
Including additional layers in the framework and/or  a larger number of neurons do not remove the oscillations.

The last neural network framework is based on convolution layers. The outputs of this framework are the $C_p$ and $C_f$ distributions.
The required error convergence property increasing the number of points in each dataset and the low sensitivity to sequence type are satisfied as depicted in the error map.
Looking closely in this regard, adding the physics-based loss further improves the performance.
Even when the uniform grid distribution is adopted the error is comparable to the other distributions, confirming the necessity of include as many physical relations as possible in the model.
The mean error in the \textit{CNN Coefficient} framework is always lower than the \textit{CNN MSE} one.
Furthermore, the error always goes down when using a larger dataset with the more sophisticated loss.
On the contrary, the \textit{CNN MSE} has a peak in the mean error for 100-point Latin hypercube and 216-point Sobol' sequences.
The Halton sequence is the one that reaches the lowest global error, but at the same time it is the one that has the largest error when using less points.
The Latin and Sobol' ones, instead, present a more limited error variation while reaching satisfying performances for 100 and 216 points.
Overall, the Latin hypercube grants better prediction quality with respect to Sobol.
To further remark the effect of physics information in the model, in Figures~\ref{img_nn_coefficients_light}~and~\ref{img_nn_coefficients_deep} it can be noted how the curves for all the three coefficients are closer to the actual solution and the differences of the datasets are less noticeable.
\begin{figure*}
	\centering
	\includegraphics[width=\textwidth]{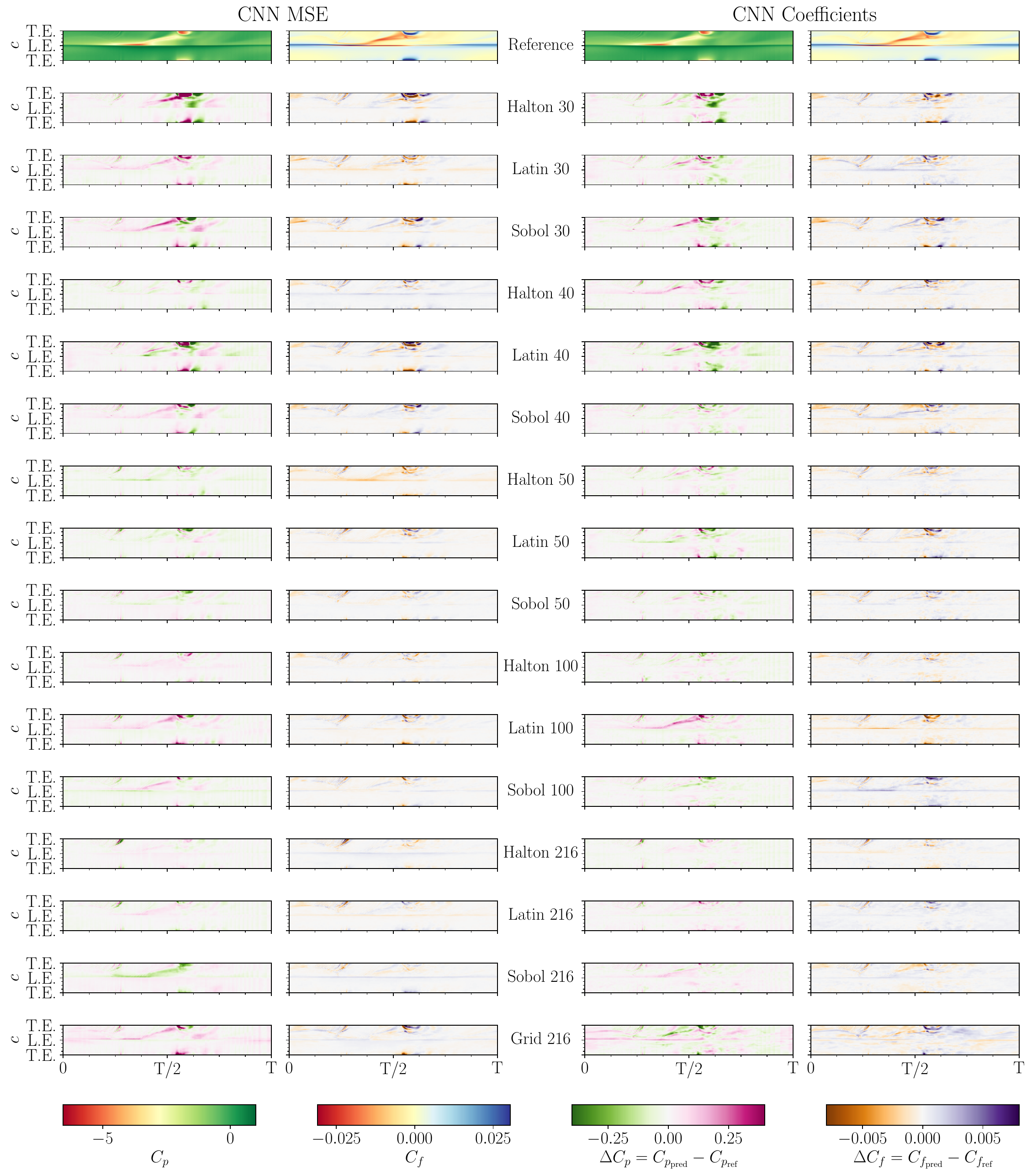}\\
	
	\caption{CNN performance comparison for a light dynamic stall example, corresponding to an oscillation amplitude of ${8.62^\circ}$, a 18.96 m/s free-stream velocity and a reduced frequency of 0.157.}
	\label{img_nn_imshow_light}
\end{figure*}
\begin{figure*}
	\centering
	\includegraphics[width=\textwidth]{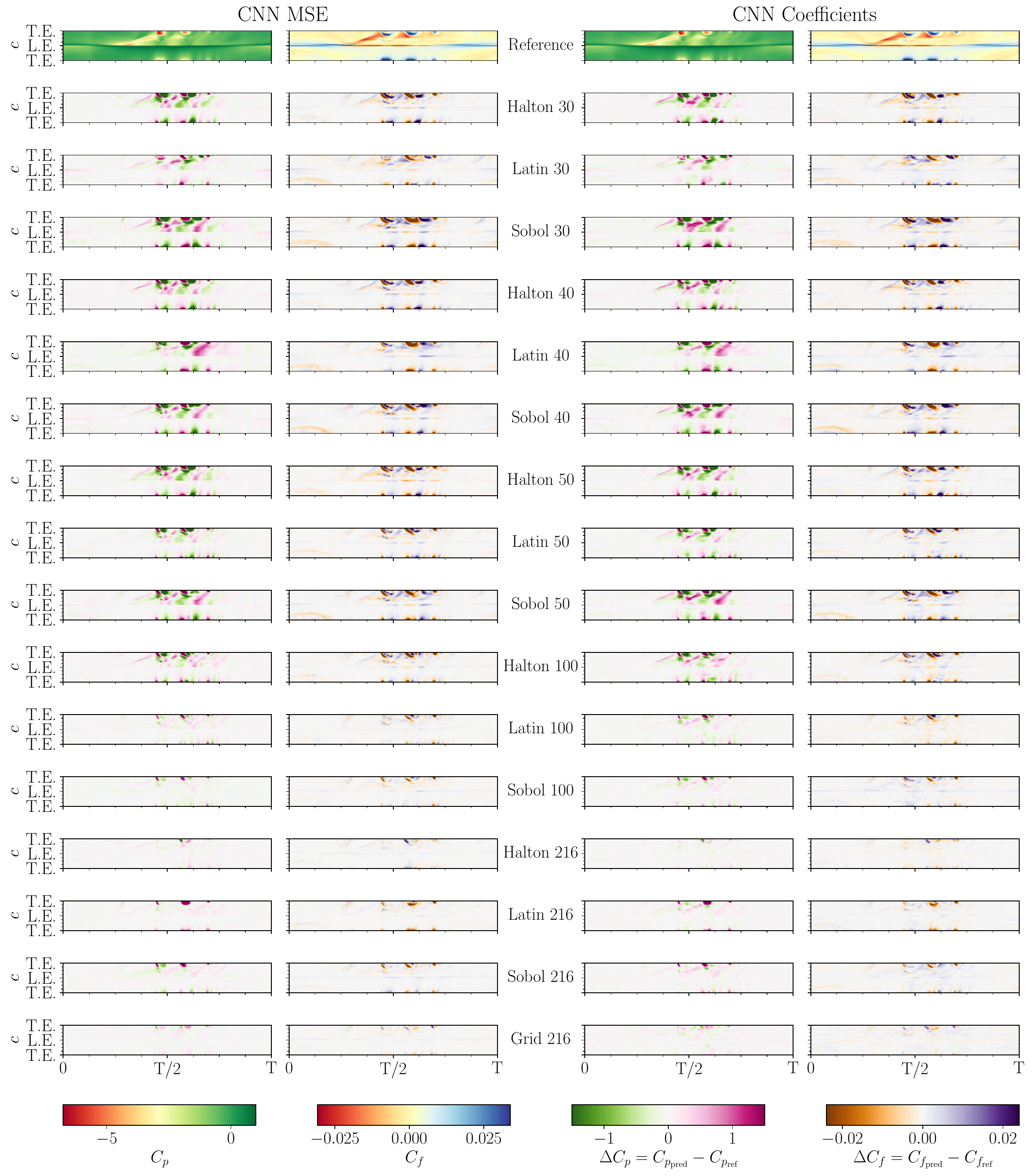}\\
	
	\caption{CNN performance comparison for a deep dynamic stall example, corresponding to an oscillation amplitude of ${14.99^\circ}$, a 15.71 m/s free-stream velocity and a reduced frequency of 0.105.}
	\label{img_nn_imshow_deep}
\end{figure*}
A final comparison is provided for the actual output of CNN based models, and it is reported in Figures~\ref{img_nn_imshow_light}~and~\ref{img_nn_imshow_deep} in which the reference $C_p$ and $C_f$ distributions are reported with the absolute error for all the employed training datasets.
Looking at the pressure and skin friction over the airfoil, the reasons of the discrepancies are highlighted. 
A common issue is the ability to capture the instabilities near the trailing edge during the upstroke phase due to the transitioning of the boundary layer from laminar to turbulent.
This behavior is mostly due to the up-sampling operations of the flow field that tends to smear out rapid oscillations.
The other region where the error is larger coincides with the pressure peaks especially for the secondary vortex that involves highly non-linear phenomena and reaches $C_p$ values close to the dynamic stall vortex that detaches first.
A peculiarity that seems in contrast to what has been presented until now is the extremely low error for the uniform grid in the deep dynamic stall case.
However, this behavior is due to another effect, that is the distance between the predicted point and the closest training point in the flight envelope.
Specifically, for this case, the closest training point is located at $\alpha=15^\circ$, $V_\infty = 16$ m/s, and $k=0.1$ which is particularly near to the deep dynamic stall test conditions.

\section{Conclusions and final remarks} \label{sec_conclusions}
This work presents a robust framework to fast predict aerodynamic loads for sinusoidal pitching airfoils incurring in light and deep dynamic stall.
A data-driven reduced order model, based on deep neural networks and high-fidelity simulations, is build to accurately describe dynamic stall phenomena.
Four different models, constructed upon three different deep learning frameworks, have been presented.
The gradual increase of complexity in the neural network frameworks has been driven by the need of a more robust prediction while maintaining the inference time as low as possible.
A physics-based loss function that incorporates the computation of lift, drag, and moment coefficients in the network has been implemented, highlighting the advantages of including physical knowledge in the model.
Moreover, a periodic condition has been implemented in the convolution layers to ensure the physical meaning of the $C_p$ and $C_f$ outputs.
Indeed, periodicity grants that there are no discontinuities in the distributions over the airfoil and also that the aerodynamic loads are continuous between subsequent cycles.
Furthermore, an extensive analysis of the training dataset point distributions has been performed.
A uniform grid distribution is compared against low discrepancy sequences: Latin hypercube, Sobol, and Halton for different dataset sizes creating a map of the average error.

The convolutional neural network, coupled with the physics-based loss function, showed unprecedented performance in predicting aerodynamic loads for a broad range of operating conditions in the prescribed flight envelope.
In addition, it has been proven that the neural network is slightly influenced by the training point distribution, even if the Latin hypercube is the one that closely predicts the reference loads.
Another fundamental feature of the model is the monotonically decreasing error, over the test datasets, when the train datasets have a larger number of points.
Despite the limitations of using three parameters to describe the flight envelope and considering incompressible CFD solutions, the properties of the presented framework are still valid for a full compressible setup, and it is a candidate to fill the gap of next-generation reduced order models for dynamic stall phenomena.

\section*{Acknowledgments}
The authors acknowledge Leonardo SpA -- Helicopter Division for guaranteeing us access to the \textit{davinci-1} supercomputer where all the computations are performed.

\end{document}